# Random Switching for High Performance DC-DC power converters

Jacques A. Naudé, Ivan W. Hofsajer, *Member, IEEE*

***Abstract*— Random Pulse Width Modulation (RPWM) has been successfully applied in power electronics for nearly 30 years. The effects of the various possible RPWM strategies on the Power Spectral Density have been thoroughly studied. Despite the effectiveness of RPWM in spreading harmonic content, an appeal is consistently made to maintain the textbook Pulse Width Modulation scheme 'on average'. Random Switching (RS) does away with this notion and probabilistically operates the switch. In addition to fulfilling several optimality conditions, including being the only viable switching strategy at the theoretical limit of performance and having lower switching losses than any other RPWM; RS allows for design of the DC behaviour separately from that of the PSD. The pulse amplitude probability affects the DC and total PSD. The first and second moment of the pulse length probability distribution affects the shape of the envelope of the noise of the PSD. The minimum pulse length acts like a selective harmonic filter. The PSD can therefore be shaped without external filtering by changing these probabilities. Gaussian and Huffman pulse length probabilities are shown to be good choices depending on whether real-time PSD control or spectrum usage are the design goal. In addition, it is shown that Cúk's state space averaging model applies to RS and FRS, with $D \to p$, hence no new tools are needed to understand the low frequency behavior or control performance. A benefit of closed loop random switching is that no filtering of the controlled variable is required. Randomly responding in a biased manner dependent on the error is hence shown to be useful. There are several good reasons to consider RS and FRS for high performance applications.**

## I. INTRODUCTION

For a DC-DC converter, Pulse Width Modulation (PWM) guarantees that in a time frame $T$, a contiguous portion of time $DT$ is dedicated to keeping a given circuit in a particular configuration. This means that, on average, $D$ of the time will be spent in this configuration [1].

Whilst PWM is a natural means of time sharing between multiple circuit configurations, there are problems concerning the additional harmonics it introduces. The additional harmonics are unwanted since the goal is DC-DC conversion. Hence, Random Pulse Width Modulation (RPWM) was invented as means to counter-act the switching harmonics from ever being produced in the first place. According to [2], the first paper describing a form of RPWM was [3] and was published in 1987. There are different names for the different types of this kind of switching: random pulse position modulation (RPPM), random carrier-frequency modulation with fixed duty cycle (RCF-FD) amongst a few others [4]–[7]. PWM consists of three parameters per pulse, namely delay, width and period. Given that each parameter may either be fixed or random, there are exactly $2^3 = 8$ possible RPWM schemes. A paper which first proposed randomizing all three PWM parameters using a Field Programmable Gate Array was completed in 2011 by Dousoky *et al* [8]. Names were also supplied for the three RPWM schemes which had not been discussed in literature before this time; these are Random Duty Ratio with RPPM and Fixed Carrier Frequency (RDRPPMFCF), Random Carrier Frequency with RPPM and Fixed Duty Cycle (RCFRPPMFD) and RCF with RPPM and Random duty ratio (RRRM) [8]. Some versions of switching remove particular harmonics through the degree of freedom they alter (for example period selection) [9]. There are sophisticated adaptive RPWM schemes which remove frequencies which resonate with the circuit's parasitics from the possible set [10]. A mixed mode controller which swaps between RPWM for EMI suppression and conventional Digital PWM to achieve good transient response was reported in [11]. Sources of randomness include Linear Feedback Shift Registers (LFSR) and analog chaotic oscillators [12]–[14]. The efficiency of RPWM was investigated and improved upon using a novel two-level random switching scheme in [15]. For high frequency switching with high control resolution, a dithered sigma-delta modulated switching was described in [16]. RPWM has even found applications in fast wireless power transmission, with superior control performance and reduced spurious emissions [17]. By making the random switching period and duty cycle dependent on each other, Kirlin *et al* allowed for selective harmonic elimination in [18]. Optimal design of randomly modulated inverters and optimal spreading of discrete harmonic power were reported in [7] and [19] respectively.

Many reviews have been written over the years [5], [20], [21]; with new results and suggestions continuously being published [22]–[27]. A few notable PhDs are Stankovic's and

Date submitted:

This work is based on the research supported in part by the National Research Foundation of South Africa (UNIQUE GRANT NO: 98251).

J. A. Naude is a PhD student at the University of the Witwatersrand, Johannesburg, South Africa (e-mail: janaude@gmail.com).

I. W. Hofsajer is head of the Future Electrical Engineering Technology research group at the University of the Witwatersrand, Johannesburg, South Africa (e-mail: ivan.hofsajer@ieee.com).

Bech's [2], [28]. Much of the theoretical framework was laid down by Stankovic in [28]. He also introduced Markov-chain driven random modulation and was awarded a patent for this work [28]–[30]. These have been explored more recently for applications [31]. Bech refined the analysis and produced substantial experimental verification of unity-gain RPWM schemes [2]. Aspects of digital closed loop control of these devices were investigated in that thesis as well [2]. An interesting application of Wold's theorem applied to random digital signals was described in [32]; whereby any stationary random process may be broken up into a predictable part and a regular random part. The implications for the power spectral density are that there is a discrete part and a continuous part, a result which Stankovic showed much earlier [28].

Given this rich history of random pulse width modulation, the present work is a deeper look at a neglected switching scheme first described (for a special parameter value) in [7]. As a generalization of the idea in [7], consider independently randomly switching to one of two possible configurations with a probability $p$. The law of large numbers shows that doing this will yield the result that, on average, $p$ of the time will be spent in this configuration [33], [34]. It will be shown in this paper that $p$ may replace $D$ in the standard circuit averaging analysis; with the low frequency behavior of the device remaining essentially unchanged when compared with conventional PWM. This class of switching strategies have been dismissed out of hand since they do not guarantee volt-second or charge balance per given time frame [2]. Whilst this is true, it will be shown that, with a very high probability, finite time volt-second or charge balance is assured. Moreover, it will be shown that volt-second and charge balance are artifacts of the principle assumption used in all DC-DC converter analysis, namely the DC plus ripple model.

A very important aspect of this kind of random switching (RS) and its generalization, Fully Random Switching (FRS), is that they both allow for complete separation of the design of the DC behavior of the circuit and the PSD. In addition, these two switching schemes offer the lowest switching losses possible out of all of the possible RPWM schemes.

The time domain control performance of RS has also not been documented before. With closed loop feedback, the extent of the ripple voltages and currents can be guaranteed whilst at the same time spreading the harmonic content as widely as is physically possible. As we approach the limits of performance of switching devices and materials, there are multiple reasons to consider these two switching strategies as superior in every way to conventional PWM in the high performance regime.

### A. *Structure of Paper and Assumptions*

It is assumed throughout that an arbitrary power converter cannot switch faster than a fundamental unit of time, $t_\epsilon$. All other switching periods are therefore integer multiples of this unit of time.

Without loss of generality, it is assumed that there are only 2 switching states and therefore only 2 circuit topologies to consider. The extension to $W$ possible circuit topologies is a small extension to the foundational work presented here.

Section II introduces the fundamental limits of control due to the primary assumption of $t_\epsilon$ being the fundamental unit of time. Section III introduces random switching and its generalization, fully random switching, and discusses probable volt-second and charge balance. Section IV develops the analytical expressions of the random switching and fully random switching power spectral densities. These are verified by Monte-Carlo simulation and an envelope approximation is also proposed which is useful for design purposes. Optimal random switching schemes are developed herein as well. Section V provides the framework used to model the time domain performance of random switching. It is proven that Cúk's state space averaging differential equation is arrived at by taking the limit towards infinitely fast switching. In addition, volt-second and charge balance are shown to be modeling artifacts of the standard DC plus ripple framework. Section VI deals with the in-circuit frequency domain aspects of random switching. A buck converter is analysed using the time domain and frequency domain methodology. Finally Section VII deals with the control of these types of devices. Quasi-static, Random Integral with State Feedback and Random switching with Hysteresis are introduced.

## II. FUNDAMENTAL LIMITS OF CONTROL

The fundamental unit of time, $t_\epsilon$ dictates the possible resolution of PWM control. This is because duty cycles have to be integer multiples of $t_\epsilon$, by definition.

By definition, the period $T$ of any conventional PWM control scheme with the above constraint is always equal to $Nt_\epsilon$ units of time, for some (possibly large) integer $N$. Hence, the smallest increment possible to any given duty cycle is $1/N$ and this represents the resolution of control, duty cycles cannot be made smaller than this value. Stated differently, given a fundamental unit of time, the duty cycle resolution and switching period are inversely proportional. Higher switching frequencies come at the cost of reduced resolution.

In the limit as $N \rightarrow 1$, which is the fastest one can possibly drive the power converter, conventional PWM is no longer possible at all (other than the trivial cases of 0% and 100% duty cycle).

Even the RPWM switching schemes suffer from this problem. For a given period, $Nt_\epsilon$ units of time, there are $2^N$ possible switching waveforms that could possibly take place. This would be $W^N$ in the case of W possible circuit configurations.

All RPWM waveforms, depending on the constraints present, are a larger or smaller subset of these $2^N$ possible switching waveforms.

For example, Random Pulse Position Modulation (RPPM) with a fixed duty cycle, $D = m/N$ would only use $N - m + 1$ of the $2^N$ possible waveforms to randomly choose from. Let the switch sequence be associated with a binary number, where each bit is the switch state at that point in the sequence. For example: an $N = 5, m = 3$ RPPM switching sequence can only be one of $11100, 01110,$ or $00111$. There are $5 - 3 +$ $1 = 3$ of these switching sequences. Mathematical induction completes the proof.

As a more extreme example, consider a RPPM with Random Carrier Frequency (RCF) modulation such that the duty cycle is fixed at $D = m/N$. This was first described in [28]. In effect, a random period is chosen, the duty cycle is fixed in terms of this and the start of the pulse is randomly chosen to be within the given period. The upper $T$ (lowest frequency) is given by $N_{max}$ and the lower $T$ (highest frequency) is $N_{min}$. It is considered that every possible $T$ in between the upper $T$ and lower $T$ is possible. The total number of possible switching signals in this family is given by $2^{N_{max}}$ since this is the longest period possible. This combination of RCF and RPPM switching schemes uses $n_{\mathrm{PF}}$ of the possible $2^{N_{max}}$ signals possible, where

$$n_{PF} = \sum_{i=0}^{N_{max}-N_{min}} (m_i - N_i + 1). \tag{1}$$

The exact solution for (1) uses the fact that, $N_i = N_{min} + i$ and $m_i = m_{min} + i$ and hence $n_{\mathrm{PF}} = (N_{max} - N_{min} + 1)(m_{min} - N_{min} + 1)$. This is far less than $2^{N_{max}}$.

What these two examples show is that the number of possible switching signals within a given time frame exponentially outweigh the signals chosen to be part of the set of possible outcomes of conventional random pulse width modulation schemes. It can be argued that this is because conventional thinking requires that some form of PWM signal must be present within a given time frame (period of repetition).

If the belief is held that it is an imperative to maintain a duty cycle, then this vast culling of signals from the set of all these possible signals will continue to occur.

## III. Random Switching

The impetus for studying this kind of switching scheme was brought about by wanting to understand the theoretical limit of DC-DC converters. Since $t_\epsilon$ is the smallest unit of time possible, the switching signal cannot have a pulse which has a length smaller than $t_\epsilon$. At this time scale, the only degree of freedom available is whether the switch is open or closed.

Formally now, consider a switching signal $q(t)$ which is either 0 or 1 at any instant in time. The transition to move between these two extremes takes no time which is to say that the derivative is not well defined at the switching transitions. Again, the switching signal is operating at the limit of possible performance. In any period of time, $kt_\epsilon$, for the duration of the minimum pulse length, the only possible state is either 1 or 0. Let this amplitude value be equal to $a_k$.

The switching function would then be described by

$$q(t) = \sum_{k=-\infty}^{\infty} a_k R_{ect}\left(\frac{t - kt_\epsilon}{t_\epsilon}\right) \tag{2}$$

where the rectangle function is defined by

$$R_{ect}(t) = \theta(t) - \theta(t-1) \tag{3}$$

and $\theta(t)$ is the Heaviside step function [35]. Hence, any switching function which operates right at the limit of what is physically possible will be described by (2). Much more though, *every possible* switching function is contained within equation (2), for various choices of $a_k$. For example, a standard PWM waveform which takes place over $N$ time frames, $m$ of which are 1's is given by $a_k = 1$ for $k$ mod $N \leq m$ and $a_k = 0$ otherwise. As another example, if every $a_k$ were chosen to be 1 at random with a probability equal to $p$, every possible switching sequence would occur with certainty as time went to infinity. Hence, this kind of Random Switching (RS) is the super-set of all possible switching schemes. Every possible switching sequence will eventually be output from this kind of scheme, though some sequences will be more frequent than others depending on $p$.

Hence, (2) is a universal description for every physically possible switching scheme.

Operating right at this theoretical limit, it is not possible to define a PWM waveform. But by using RS, which is formally defined by the following probability distribution,

$$\begin{aligned} \mathbb{P}\{a_k = 1\} &= p \\ \mathbb{P}\{a_k = 0\} &= 1 - p = p', \end{aligned} \tag{4}$$

the time average of the expected value can be maintained. To see this, note that the probability of $a_k$ is independent of the time-frame under consideration $k$. This gives it the special property that it may be shifted outside of the angle bracket operator (see Appendix). Hence,

$$\langle \mathbb{E}\{q(t)\} \rangle = \mathbb{E}\{a_k\} \left\langle \sum_{k=-\infty}^{\infty} R_{ect}\left(\frac{t - kt_\epsilon}{t_\epsilon}\right) \right\rangle \tag{5}$$

Since the term inside the angle bracket is periodic, the angle bracket is equal to the time average over a single period of one of the rectangular pulses. The time average of the rectangular pulse is unity and hence

$$\langle \mathbb{E}\{q(t)\} \rangle = \mathbb{E}\{a_k\} = p \times 1 + p' \times 0 = p \tag{6}$$

Therefore, by randomly switching the amplitudes independently with probability distribution given by (4) a measure of control of the average of the switching sequence is given by $p$, the probability of switching the amplitude to 1.

As a comparison, taking the angle bracket of the expected value of a deterministic PWM signal can be shown to be equal to $D$. Hence, $p$ fulfills the same role as $D$ does in a conventional PWM switching scheme.

### A. *Probable Volt-Second or Charge Balance of RS*

Consider a PWM converter operating at steady state. It is well known that if the fractional "on" time in a given period is equal to the appropriate duty cycle for that steady state, then volt-second or charge balance will be assured for the state variables [2].

All presently used RPWM switching schemes have contiguous blocks of "on" time in a single period due to the fact that it is a PWM signal which is being randomly perturbed. Those which maintain volt-second or charge balance on a per period basis have a fixed duty cycle for each single period.

There is no reason for volt-second or charge balance to be disrupted if this "on" time were broken up into smaller time intervals and shuffled within this single period. The time exposure of the state variable to the two possible rates of change is the same in both the contiguous and the shuffled version (assuming linear ripple). Hence, volt-second or charge balance will be assured for the shuffled version as well. For example, $0011100$ would have the same "on" time as $0101010$, they both have three "on" states in a sequence that is 7 units of time long.

The conclusion is therefore the following: under linear ripple and at steady state, if the fractional "on" time in a given period is equal to the appropriate duty cycle, then volt-second or charge balance is assured.

Now consider the behavior of RS. Given a finite time-frame $Nt_\epsilon$, the total number of 1's that occur is a random variable, $n_1$.

This random variable is fully characterised by the binomial distribution

$$n_1 = \binom{N}{n_1} p^{n_1}(1-p)^{N-n_1} \tag{7}$$

since this is the same result as counting the number of heads that will occur during the flipping of a coin with probability of returning heads equal to $p$ [36]–[38].

The expected number of 1's can be found using the standard generating function methodology [39]. The results are $\mathbb{E}\{n_1\} = Np$ with the variance given by $\mathbb{V}\{n_1\} = Npp'$. The expected fractional "on-time" is therefore calculated by $\mathbb{E}\{n_1 t_\epsilon / Nt\epsilon\} = p$ with a variance given by $\mathbb{V}\{n_1 t_\epsilon / Nt_\epsilon\} = Npp'/N^2 = pp'/N$. Hence, under RS, an estimate of the fractional "on-time" or duty cycle in a length of time $Nt_\epsilon$ is equal to

$$D_N = \mathbb{E}\left\{\frac{n_1}{N}\right\} \pm \sqrt{\mathbb{V}\left\{\frac{n_1}{N}\right\}} = p \pm \sqrt{\frac{p(1-p)}{N}} \tag{8}$$

This is what was meant by probable volt-second or charge balance. For any given time frame $Nt_\epsilon$ units long, RS has an error in volt-second or charge balance on the order of $N^{-1/2}$.

As the time frame considered is extended, the probability of volt-second or charge balance asymptotically approaches certainty.

Further on in this paper, it will be shown that Volt-second and charge balance are modelling artifacts. Volt-second and charge balance is hence assured by definition.

## B. Switching Losses in RS

The usual model for switching losses in hard switching is well known [1]. The energy lost per switching event depends explicitly on the switch realization and the materials used for switching. The number of switching transitions per period is a proxy for the switching losses since it is only during transitions that energy is lost (if ZVS or ZCS is not used). Conduction losses are normally accounted for separately.

Every existing RPWM scheme which does not include duty cycles of 0% and 100% in the set of admissible values has a guaranteed switching transition up and a guaranteed switching transition down in every possible period. Using these facts, it is not difficult to prove for all RPWM which do not include 0% and 100% duty cycles that

$$\langle \mathbb{E}\{P_{loss}\}\rangle = \frac{W_{on} + W_{off}}{\bar{\tau}} = \left(W_{on} + W_{off}\right)\overline{f_s} \tag{9}$$

where $\bar{\tau}$ is the average period in the RPWM set (and $\overline{f_s}$ is the average switching frequency); $W_{\text{on}}$ is the turn-on energy loss and $W_{\text{off}}$ is the turn-off energy loss.

With RS and FRS, the above is no longer true. It may be that a pulse with an amplitude of 1 is followed by another pulse with amplitude 1. In this case the switch does not change state and there is no switching transition to realise an energy loss. Using the law of large numbers, for RS and FRS the average expected switching loss is given by

$$\langle \mathbb{E}\{P_{loss}\}\rangle = \left(W_{on}\langle \mathbb{E}\{n_{on}\}\rangle + W_{off}\langle \mathbb{E}\{n_{off}\}\rangle\right), \tag{10}$$

where $\langle \mathbb{E}\{n_{\text{on}}\}\rangle$ is the average expected number of turn-on transitions in an average period. Note that in any RPWM and PWM, $\langle \mathbb{E}\{n_{\text{on}}\}\rangle = \langle \mathbb{E}\{n_{\text{off}}\}\rangle = 1$, since it is guaranteed that there is a single turn-on event and a single turn-off event; hence (9) is recovered.

Counting the number of transitions in this random switching scheme is not trivial. It is most easily accomplished by taking the time derivative of the switching function and analyzing the weights of the Dirac deltas which result from the taking the derivatives of the Heaviside step functions that make up the rectangular pulses. A Dirac delta with a weight of +1 is a turn-on transition and a weight of -1 is a turn-off transition.

For RS the derivative of the switching function is given by

$$\frac{dq}{dt} = \sum_{k=-\infty}^{\infty} a_k\left(\delta(t - kt_\ell) - \delta(t-(k+1)t_\ell)\right) \tag{11}$$

where $t_\ell = \ell t_\epsilon$. It is now possible to combine the weights of the Dirac deltas in (11) to produce

$$\frac{dq}{dt} = \sum_{k=-\infty}^{\infty} (a_k - a_{k-1})\delta(t - kt_\ell). \tag{12}$$

A very important property of the absolute value applied to (12) is that

$$\left|\frac{dq}{dt}\right| = \sum_{k=-\infty}^{\infty} |a_k - a_{k-1}|\delta(t - kt_\ell). \tag{13}$$

Of course, the absolute value is not linear and the absolute value of the sum is not equal to the sum of absolute values in general. However, the absolute value of the derivative of the switching function is zero wherever there is no Dirac delta present, so the total sum is zero wherever $t \neq kt_\ell$; hence the absolute value may be placed within the summation.

Furthermore $|w\delta(t)| = |w|\delta(t)$ since the absolute value of the Dirac delta is undefined; it is its weight which matters.

Hence, using these considerations, the absolute value equation in (13) is correct. Since the absolute value of the weights now label that a transition has occurred (and not whether it was up or down), it is possible to count the average expected number of transitions by finding the average expected value of (13). That is $\langle \mathbb{E}\{n_{\text{tot}}\}\rangle = \langle \mathbb{E}\{|\dot{q}|\}\rangle$, where $n_{\text{tot}}$ is the total number of switching transitions in an average period.

Counting the average expected number of turn-on transitions can be calculated with,

$$\left\langle \mathbb{E}\left\{\left|\frac{dq}{dt} + 1\right| - 1\right\}\right\rangle = \frac{\mathbb{E}\{|a_k - a_{k-1} + 1| - 1\}}{t_\ell} = \frac{p(1-p)}{t_\ell} \tag{14}$$

since the $+1$ in the absolute value zeroes out the turn-off transition weights and the $-1$ outside of the absolute value removes this introduced bias. It is important to note also that $\langle \sum_k \delta(t - kt_\ell)\rangle = t_\ell^{-1}$; which can be shown from the definition of the angle bracket, rewriting the limit in terms of multiples of the pulse length and the definition of the integral of the Dirac delta. The reader can convince themselves of the validity of (14) by working through every possible pulse amplitude (there are only 4) and noting that the only non-zero value $|a_k - a_{k-1} + 1| - 1$ is equal to 1. This occurs with probability $\mathbb{P}\{a_k = 1, a_{k-1} = 0\} = p(1-p)$ since these are independent and identically distributed amplitude probabilities. Similarly, the average expected number of turn-off transients per average period are calculated by looking at $\left\langle \mathbb{E}\left\{\left|1 - \frac{dq}{dt}\right| - 1\right\}\right\rangle$. This can be calculated from first principles and be shown to be equal (14) or one can use the fact the turn-on transitions must have a matching number of turn-off transitions, on average. Hence, the switching losses of both RS and FRS can be shown to be equal to

$$\langle \mathbb{E}\{P_{loss}\}\rangle = \left(W_{on} + W_{off}\right)p(1-p)\overline{f_s} \tag{15}$$

where $\overline{f_s} = \langle \mathbb{E}\{t_\ell^{-1}\}\rangle$ and is the average switching frequency. Note that $p(1-p)$ is always smaller than 1 and therefore RS and FRS will always have a smaller average expected switching loss than any conventional RPWM scheme, as evidenced by comparing (9) and (15). It is therefore proven that for the same average switching frequency, RS and FRS are more efficient than RPWM and PWM since the proposed random switching schemes both have reduced switching losses calculated by (15).

## IV. Spectral Analysis of RS

As is customary when analysing the power spectrum of various switching strategies, it is the power spectrum of the *switching signal* which will be analysed. The power spectrum of the actual circuit variables will analysed in the sequel as these are deterministic functions of the switch and can be inferred after some manipulation [28].

### *A. Power Spectrum of the RS Function*

Calculating the spectral content of a random process is well known and regularly applied in the power electronics literature [2], [5], [28], [30], [33], [40]. The most important fact is that the Power Spectral Density (PSD) of a randomly switched signal is *not* just the Fourier transform of it. Furthermore, it is important to note that

$$\int_{-\infty}^{\infty} \mathcal{S}_{qq}(f)df = \langle \mathbb{E}\{q(t)^2\}\rangle \tag{16}$$

which is the Parseval-Plancherel theorem for random processes [33]. This result is important because it acts as a conservation law of PSD in cases where the average expected square values are the same. Hence, even if the underlying switching regimes are different, provided that two switching schemes have the same average expected squared value, their total PSD will the same. The difference will be in the distribution amongst the frequencies of the PSD in each case.

In the present case,

$$\langle \mathbb{E}\{q(t)^2\}\rangle = \left\langle \sum_k \mathbb{E}\{a_k^2\} R_{ect}\left(\frac{t - kt_\epsilon}{t_\epsilon}\right)^2 \right\rangle \tag{17}$$

because only when $k = n$ is the product of the rectangle functions non-zero. The expected amplitude squared may be taken outside of the angle bracket now. Hence,

$$\begin{aligned} \langle \mathbb{E}\{q(t)^2\}\rangle &= \mathbb{E}\{a_k^2\}\frac{1}{t_\epsilon}\int_0^{t_\epsilon} R_{ect}\left(\frac{t}{t_\epsilon}\right)^2 dt \\ &= \left(p \times 1^2 + (1-p) \times 0^2\right)\frac{1}{t_\epsilon}t_\epsilon = p \\ &\Rightarrow \int_{-\infty}^{\infty} \mathcal{S}_{qq}(f)df = p. \end{aligned} \tag{18}$$

This value is therefore the total PSD and is a useful check for the calculations which follow. This is an important result. It was shown that $p$ is the average expected value of the switching signal in (6), here it is now shown that the total PSD of the switching function is also equal to $p$. For standard PWM, by definition the average "on" time is equal to $D$ and it can be shown that the total PSD is also equal to $D$.

This is not a coincidence, *all* switching signals which have the same average expected value, $\langle \mathbb{E}\{q(t)\}\rangle$, will also have the same average expected squared value, $\langle \mathbb{E}\{q(t)^2\}\rangle$. This can be proven using the Bhatia-Davis inequality for the variance and the amplitude distribution of the switching signal [41], [42].

Hence, the sum of the squares of the all of the harmonics in a PWM switching scheme yields the same value as the integral over all frequencies of the PSD of a RS scheme provided that $p = D$. The PSD of PWM is well known. What follows is the analytical PSD of RS.

One of the simplest methods for calculating a PSD for random processes is found in [33]. The expected value of the modulus squared of the Fourier transform of a windowed version of the random signal is taken and the limit as the window goes to infinity is discovered. Explicitly,

$$\mathcal{S}_{qq}(f) = \lim_{T\to\infty} \frac{\mathbb{E}\{|Q_T(f)|^2\}}{2T} \tag{19}$$

where $Q_T(f) = \mathcal{F}\{q_T(t)\}$ and hence $|Q_T(f)|^2 = Q_T(f)Q_T(f)^*$.

Truncating the switching signal in terms of the fundamental unit of time means that $2T = (2N+1)t_\epsilon$ with the truncated switching signal given by

$$q_T(t) = \sum_{k=-N}^{N} a_k R_{ect}\left(\frac{t - kt_\epsilon}{t_\epsilon}\right). \tag{20}$$

Let the Fourier transform of an individual rectangular function be defined with $U(f)$ which is equal to

$$U(f) \coloneqq \mathcal{F}\{R_{ect}(t)\} = \frac{\mathbb{j}}{2\pi f}\left(e^{-\mathbb{j}2\pi f} - 1\right). \tag{21}$$

Hence,

$$Q_T(f) = t_\epsilon U(ft_\epsilon) \sum_{k=-N}^{N} a_k e^{-\mathbb{j}2\pi f k t_\epsilon} \tag{22}$$

and therefore

$$|Q_T(f)|^2 = t_\epsilon^2 |U(ft_\epsilon)|^2 \sum_{k=-N}^{N} \sum_{m=-N}^{N} a_k a_m e^{-\mathbb{j}2\pi f(k-m)t_\epsilon} \tag{23}$$

Now taking the expected value of (23) leads to

$$\mathbb{E}\{|Q_T(f)|^2\} = t_\epsilon^2 |U(ft_\epsilon)|^2 (2N+1)\mathbb{E}\{a_k^2\} + t_\epsilon^2 |U(ft_\epsilon)|^2 CT \tag{24}$$

This is because there are $2N+1$ terms where $k = m$ and CT is defined as the sum involving the cross terms where $k \neq m$. The symbol $\delta_{k,m}$ denotes the Kronecker delta [39], [43]. Using the Kronecker delta, the closed-form sum of the cross terms is given by

$$CT = \mathbb{E}\{a_k a_m\} \sum_{k,m} \left(1 - \delta_{k,m}\right) e^{-\mathbb{j}2\pi f(k-m)t_\epsilon} \tag{25}$$

The $(1 - \delta_{k,m})$ in (25), makes sure to exclude the $k = m$ terms from being double counted. Given the time invariance and independence of the probability of the amplitudes; $\mathbb{E}\{a_k a_m\}$ is statistically the same value for all $k$ and $m$, provided that $k \neq m$, and hence can be factored out of the sum.

By inspection for small values of $N$, and mathematical induction, the cross term sum simplifies to

$$CT = \mathbb{E}\{a_k a_m\} \sum_{\substack{n=-2N \\ n\neq 0}}^{2N} (2N + 1 - |n|)\, e^{\mathbb{j}n2\pi f t_\epsilon} - \mathbb{E}\{a_k a_m\}(2N+1) \tag{26}$$

where the last term has been computed using the sifting property of the Kronecker delta.

The ensemble average of $a_k^2$ was already calculated in (18) to be equal to $p$. Similarly it can be shown that $\mathbb{E}\{a_k a_m\} = p^2$. The modulus squared of the pulse function is given by

$$t_\epsilon^2 |U(ft_\epsilon)|^2 = \frac{sin^2(\pi f t_\epsilon)}{(\pi f)^2}. \tag{27}$$

Taking the limit $T \to \infty$ is the same as taking the limit as $N \to \infty$ when done in integer multiples of the fundamental unit of time. Hence, terms multiplied by $(2N+1)$ survive the limiting process whereas all the other terms go to zero in this case. Without further refinement, the PSD of the RS function is given by

$$\mathcal{S}_{qq}(f) = \frac{1}{t_\epsilon} t_\epsilon^2 |U(ft_\epsilon)|^2 p(1-p) + \frac{1}{t_\epsilon} t_\epsilon^2 |U(ft_\epsilon)|^2 p^2 \sum_{\substack{n=-\infty \\ n\neq 0}}^{\infty} e^{\mathbb{j}2\pi f n t_\epsilon}. \tag{28}$$

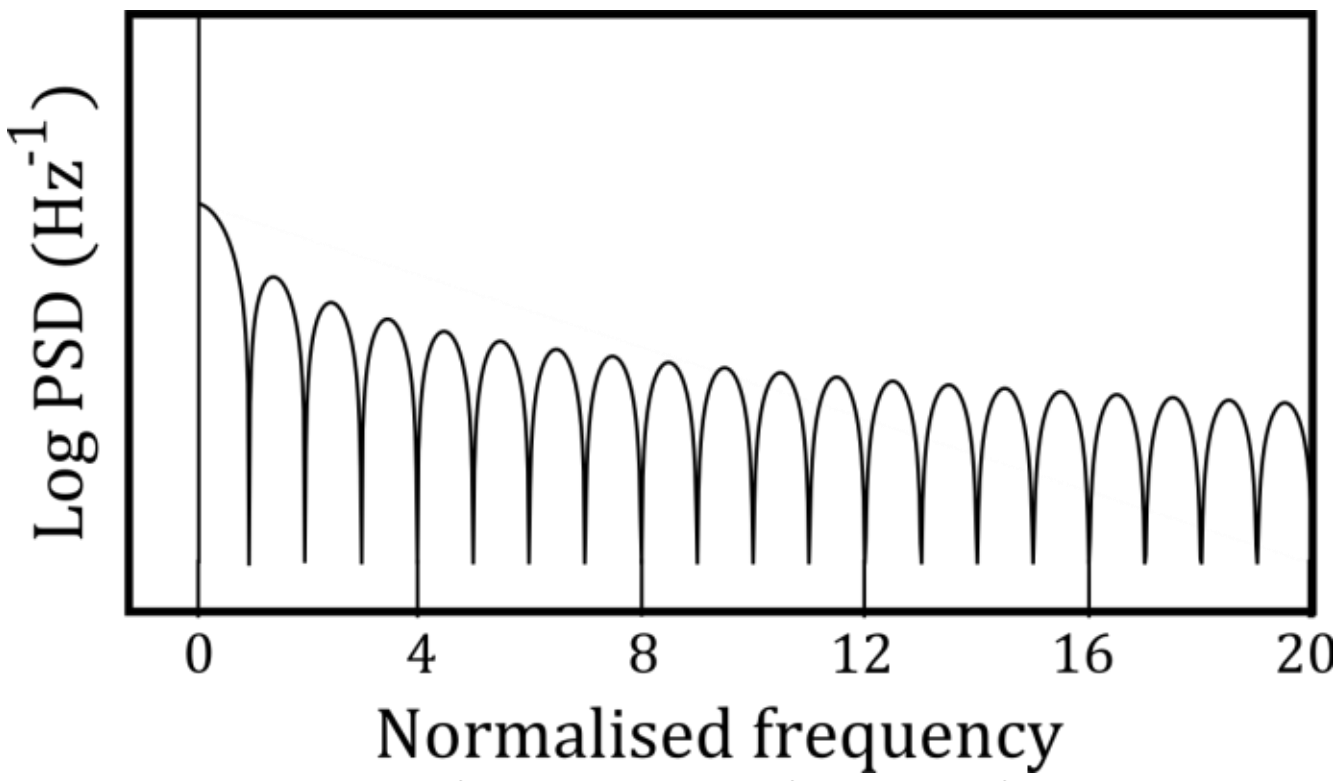

Fig. 1: Analytical PSD of RS at the limit of switching frequency. The DC component is the squared weight of the impulse at the origin. The switching noise is the sinc-like function.

The sum of complex exponentials can be simplified to be represented as a Dirac comb [2], [44]. Hence another way of representing the PSD is

$$\mathcal{S}_{qq}(f) = t_\epsilon |U(ft_\epsilon)|^2 p(1-p) + t_\epsilon |U(ft_\epsilon)|^2 \frac{p^2}{t_\epsilon} \sum_{n=-\infty}^{\infty} \delta\left(f - \frac{n}{t_\epsilon}\right) \tag{29}$$

Since $|U(ft_\epsilon)|^2$ is zero exactly where the Dirac comb is active (except at $n = 0$), the weights of all of the discrete harmonics in the Dirac comb are all zero.

Finally, the PSD of the switching function is shown to be completely continuous (except at DC) and is equal to

$$\mathcal{S}_{qq}(f) = p(1-p)\frac{sin^2(\pi f t_\epsilon)}{t_\epsilon(\pi f)^2} + p^2\delta(f) \tag{30}$$

where $\lim_{f\to 0} t_\epsilon^2 |U(ft_\epsilon)|^2 = t_\epsilon^2$ and the sifting property of the Dirac delta has been used.

If desired, the PSD of a RS scheme with a pulse length which is a multiple of $t_\epsilon$ may be found using (30) and formally substituting $t_\epsilon \to t_\ell = \ell t_\epsilon$, where $\ell$ is an integer.

As a check, $\int_{-\infty}^{\infty} \mathcal{S}_{qq}(f)\,\mathrm{d}f = p(1-p) + p^2 = p$ as it should.

The coefficient of $\delta(f)$ is the DC value of the switching signal squared and the coefficient of the sinc-like function is the total noise harmonic content of the switching signal, also known as the variance [33]. This is because every PSD can be represented as

$$\mathcal{S}_{qq}(f) = \langle \mathbb{V}\{q\}\rangle g(f) + \langle \mathbb{E}\{q\}\rangle^2 \delta(f) \tag{31}$$

where $\langle \mathbb{V}\{q\}\rangle$ is the average variance of the switching signal and $g(f)$ is the frequency dependent function which models how the spectral density varies as a function of frequency; it also integrates to unity [33]. For want of a better name, (31) can be called the universal power spectral density model.

A depiction of the logarithm of the PSD is in Fig. 1.

It is important to note that this is an optimum switching scheme if the goal is to eliminate all discrete harmonics, as would be desirable in a DC-DC converter. Stated another way, this is the most "spread out" the PSD can be without randomly altering the pulse lengths of the RS signal as well. None of the variance is concentrated at any one harmonic as would be the case with PWM or other RPWM which have discrete components.

### B. Fully Random Switching

RS has the property that only the amplitudes of the universal switching function are randomly altered. By independently randomly selecting the time that the random amplitude pulse exists for (i.e. the pulse length), the only two degrees of freedom of the universal switching function will have been randomised. Essentially $t_\epsilon$ is formally replaced by another random variable $t_\epsilon \to \ell_k t_\epsilon$. Here $\ell_k$ represents the length of the pulse in the $k$'th unit of time and is a random integer between $\min \ell$ and $\max \ell$. It is taken for granted that $\min \ell = 1$ in all cases henceforth.

This type of switching scheme, named Full Random Switching (FRS) is, in effect, RCF combined with RS. It is a switching scheme which completely utilizes both of the two degrees of freedom present in the universal switching function from (2), hence the name.

The FRS signal is represented by

$$q(t) = \sum_{k=-\infty}^{\infty} a_k R_{ect}\left(\frac{t - T_k}{\ell_k t_\epsilon}\right). \tag{32}$$

The time delay $T_k$ is specified by (33) in exactly the same way as RCF [2].

$$T_k = t_\epsilon \sum_{i=1}^{k-1} \ell_i \;\; if\; k > 0$$
$$T_k = 0 \; if \; k = 0 \tag{33}$$
$$T_k = -t_\epsilon \sum_{i=1}^{k-1} \ell_{-i} \;\; if\; k < 0$$

Note that there are now three random variables in (32), the amplitude $a_k$, a particular pulse length at $k$, $\ell_k$, and the sum of all previous pulse lengths up to the start time of a particular pulse $T_k$.

The Fourier transform of a truncated version of the FRS signal is given by

$$\mathcal{Q}_T(f) = \sum_{k=-N}^{N} a_k U_{\ell_k}(f) e^{-\mathbb{j}\omega T_k} \tag{34}$$

where $\omega = 2\pi f$ and $U_{\ell_k}(f) \coloneqq \mathcal{F}\left\{\mathrm{R_{ect}}\left(\frac{t}{\ell_k t_\epsilon}\right)\right\}$.

Calculating the magnitude squared of this function means

$$|\mathcal{Q}_T|^2 = \sum_{k=-N}^{N}\sum_{m=-N}^{N} a_k a_m U_{\ell_k}(f) U_{\ell_k}(f)^* e^{-\mathbb{j}\omega T_k} e^{\mathbb{j}\omega T_m} \tag{35}$$

The expected value is calculated by using the time invariance and independence of the amplitudes and the individual pulse lengths.

$$\mathbb{E}\{|\mathcal{Q}_T|^2\} = (2N+1)p\mathbb{E}_\ell\{|U_\ell(f)|^2\} + p^2 \mathbb{E}_\ell\{U_\ell\}\mathbb{E}_\ell\{U_\ell^*\}\sum_{k,m}(1-\delta_{k,m})\mathbb{E}\left\{e^{\mathbb{j}\omega\ell t_\epsilon}\right\}^{m-k} \tag{36}$$

where again the Kronecker delta has been used to sort out the $k = m$ terms first and subtract off the false ones from the general double sum. Note that the following properties were used to get the exponential term to be summed,

$$\begin{aligned}
\mathbb{E}\left\{e^{-\mathbb{j}\omega T_k}\right\} &= \mathbb{E}\left\{e^{-\mathbb{j}\omega(\ell_1+\ell_2+\cdots\ell_{k-1})t_\epsilon}\right\} \\
&= \mathbb{E}\left\{e^{-\mathbb{j}\omega\ell_1 t_\epsilon} e^{-\mathbb{j}\omega\ell_2 t_\epsilon} \ldots e^{-\mathbb{j}\omega\ell_{k-1} t_\epsilon}\right\} \\
&= \mathbb{E}\left\{e^{-\mathbb{j}\omega\ell t_\epsilon}\right\}^{k-1} \\
&= \mathbb{E}\left\{e^{\mathbb{j}\omega\ell t_\epsilon}\right\}\mathbb{E}\left\{e^{-\mathbb{j}\omega\ell t_\epsilon}\right\}^{k} \\
&= \mathbb{E}\left\{e^{\mathbb{j}\omega\ell t_\epsilon}\right\}\mathbb{E}\left\{e^{\mathbb{j}\omega\ell t_\epsilon}\right\}^{-k}.
\end{aligned} \tag{37}$$

Before taking the limit, the double sum can be simplified in exactly the same manner as (25). Hence, the expected value is

$$\mathbb{E}\{|\mathcal{Q}_T|^2\} = (2N+1)p\mathbb{E}_\ell\{|U_\ell(f)|^2\} \tag{38}$$

$$-(2N+1)p^2\mathbb{E}_\ell\{U_\ell\}\mathbb{E}_\ell\{U_\ell^*\}$$
$$+p^2\mathbb{E}_\ell\{U_\ell\}\mathbb{E}_\ell\{U_\ell^*\}\times$$
$$\sum_{\substack{n=-2N\\n\neq 0}}^{2N}(2N+1-|n|)\mathbb{E}_\ell\{e^{\mathbb{j}\omega\ell t_\epsilon}\}^n$$

where the limit as $T\to\infty$ is the same as the limit of $N\to\infty$ with $2T=(2N+1)\,\bar{\ell}t_\epsilon$. Note that $\bar{\ell}=\mathbb{E}\{\ell\}$ and is the expected value of $\ell$.

Hence, the PSD of the FRS signal is given by

$$\mathcal{S}_{qq}(f)=\frac{p}{\bar{\ell}t_\epsilon}\mathbb{E}_\ell\{|U_\ell(f)|^2\}$$
$$-\frac{p^2}{\bar{\ell}t_\epsilon}\mathbb{E}_\ell\{U_\ell(f)\}\mathbb{E}_\ell\{U_\ell(f)^*\} \tag{39}$$
$$+\frac{p^2}{\bar{\ell}t_\epsilon}\mathbb{E}_\ell\{U_\ell(f)\}\mathbb{E}_\ell\{U_\ell(f)^*\}\sum_{\substack{n=-\infty\\n\neq 0}}^{\infty}\mathbb{E}_\ell\{e^{\mathbb{j}\omega\ell t_\epsilon}\}^n$$

As a test for correctness, if $\ell$ becomes deterministic, $\bar{\ell}=\ell$ and all expectations with respect to $\ell$ become just that single term. Under this condition, the PSD of the FRS scheme in (39) becomes the PSD of the RS scheme in (30) as it should.

Simplifying (39) is challenging without using Parseval-Plancherel's theorem and the universal power spectral density model in (31). A sketch of the lengthy derivation of the final simplification is given by the following. Expand the term $\frac{p^2}{\bar{\ell}t_\epsilon}\mathbb{E}_\ell\{U_\ell(f)\}\mathbb{E}_\ell\{U_\ell(f)^*\}$; this yields $\frac{p^2}{\bar{\ell}t_\epsilon}\mathbb{E}_\ell\{|U_\ell(f)|^2\}+\cdots$, where ellipses denote all of the cross probability terms from multiplying out the expected values. To see this, multiply out $(p_1x_1+p_2x_2+\cdots+p_Mx_M)(p_1y_1+p_2y_2+\cdots+p_My_M)$ and consider the analogy.

At this stage of a similar calculation with RCF, Bech split the infinite sum of complex exponentials in (39) into two and used the geometric sum approximation on each [2]. This became known as Bech's approximation [18]. In our case we are very fortunate. By finding the $\mathbb{E}_\ell\{|U_\ell(f)|^2\}$ term hidden in the product of expected values the result is

$$\mathcal{S}_{qq}(f)=\frac{p(1-p)}{\bar{\ell}t_\epsilon}\mathbb{E}_\ell\{|U_\ell(f)|^2\}+\cdots + p^2\delta(f) \tag{40}$$

where the ellipses denote a complicated expression and the necessary DC term from the universal power spectral density model in (31) has been added in. Finding the total power spectral density of (106), $\int_{-\infty}^{\infty}\mathcal{S}_{qq}(f)\mathrm{d}f=p(1-p)+z+p^2=p+z$, where $z$ is the total power spectral density of the complicated expression. But the Parseval-Plancherel theorem says that this total power spectral density must be equal to $p$, therefore $z=0$. All that this shows so far is that the total power spectral density of the complicated expression must integrate to zero. To show that the complicated expression *is* zero uses the fact that any power spectral density of a real random process must be positive or zero for every frequency [33], [40]. The only way to integrate a function to zero in this case is to have that function *be* zero at all frequencies. Hence, the correct final form of the PSD of the FRS is given by (41), which is particularly simple.

$$\mathcal{S}_{qq}(f)=\frac{p(1-p)}{\bar{\ell}t_\epsilon}\mathbb{E}_\ell\{|U_\ell(f)|^2\}+p^2\delta(f) \tag{41}$$

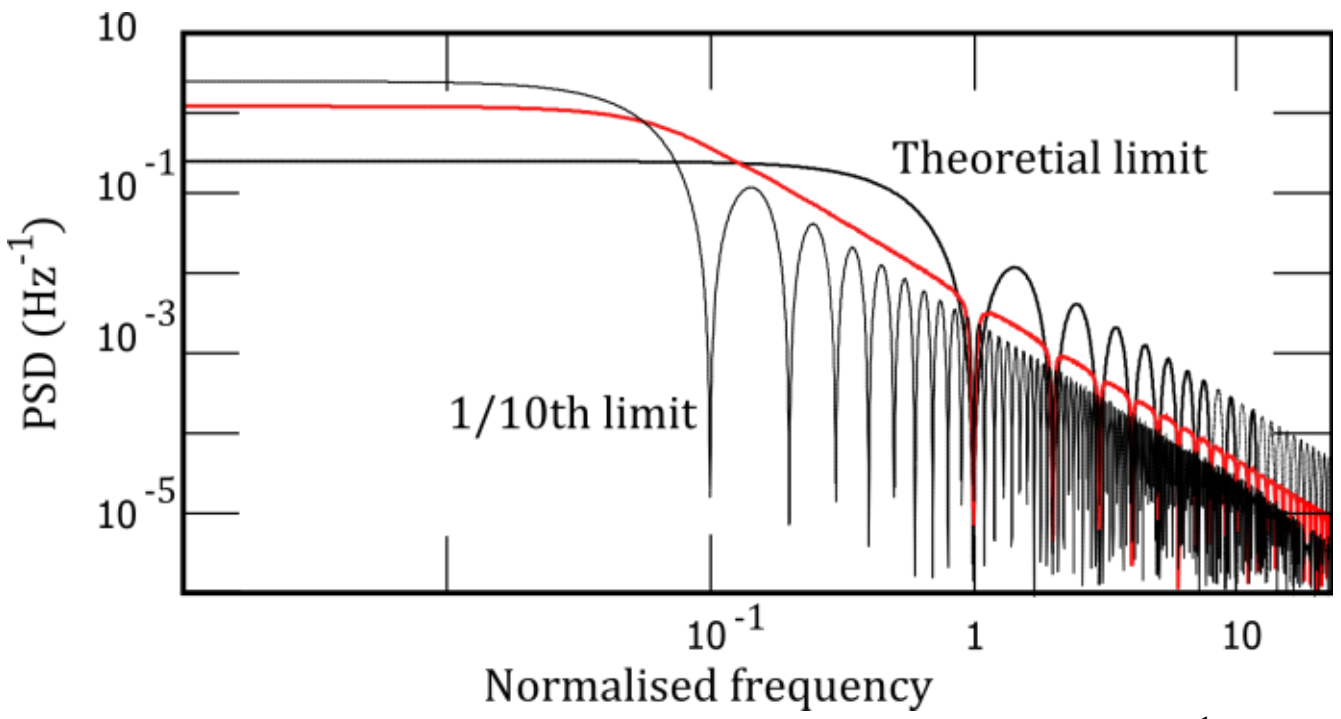

Fig. 2: RS PSD with $\ell=1$ and $\ell=10$ (black) and FRS $\mathbb{P}\{\ell\}=\frac{1}{10}$ (red). Note how the FRS PSD interpolates between the two extremal RS scheme power spectral densities. The theoretical limit has $\ell=1$.

### *C. Analytical Comparison of Random and Fully Random Power Spectral Densities*

It is now possible to compare RS and FRS in terms of their respective PSDs. If the purpose of these schemes is to achieve DC-DC conversion, then the probability of the amplitude, $p$, will be fixed by the requirements of the converter. Observe that neither (30) nor (41) have the probability of the amplitude affecting the noise PSD, other than as an overall gain. What this means is that both RS and FRS can specify the DC behavior independently from the noise distribution of the PSD. In addition, there are no discrete harmonics present at all, which is a major benefit. Therefore in both RS and FRS, selecting $p$ sets the DC value and the total PSD; whereas selecting $\ell$, the length of the pulse(s), affects the shape of the noise.

Consider operating at the very limit of possible performance, which means operating at a frequency $f_\epsilon=1/t_\epsilon$. Only RS is possible at this frequency and its analytical form is depicted in Fig. 2. Now consider switching at 1/10 of this limit ($t_\epsilon\to 10t_\epsilon$), which is also depicted in Fig. 2. Observe, that with RS, there is an inherent gain-band-width trade-off, depending on the frequency of switching.

The lower frequency RS has a lower band-width (and hence begins rolling off at a lower frequency) at the cost of an increased low frequency noise level ("gain"). The RS operating at the theoretical limit has the lowest possible low frequency noise level at the cost of rolling off at a higher frequency. This is because of the conservation law following (16). Since $p$ is the same in all cases here, the total amount of PSD is the same. Its distribution in the frequency domain is different though and this is solely affected by the pulse length, the amplitude probability does not change the distribution of noise.

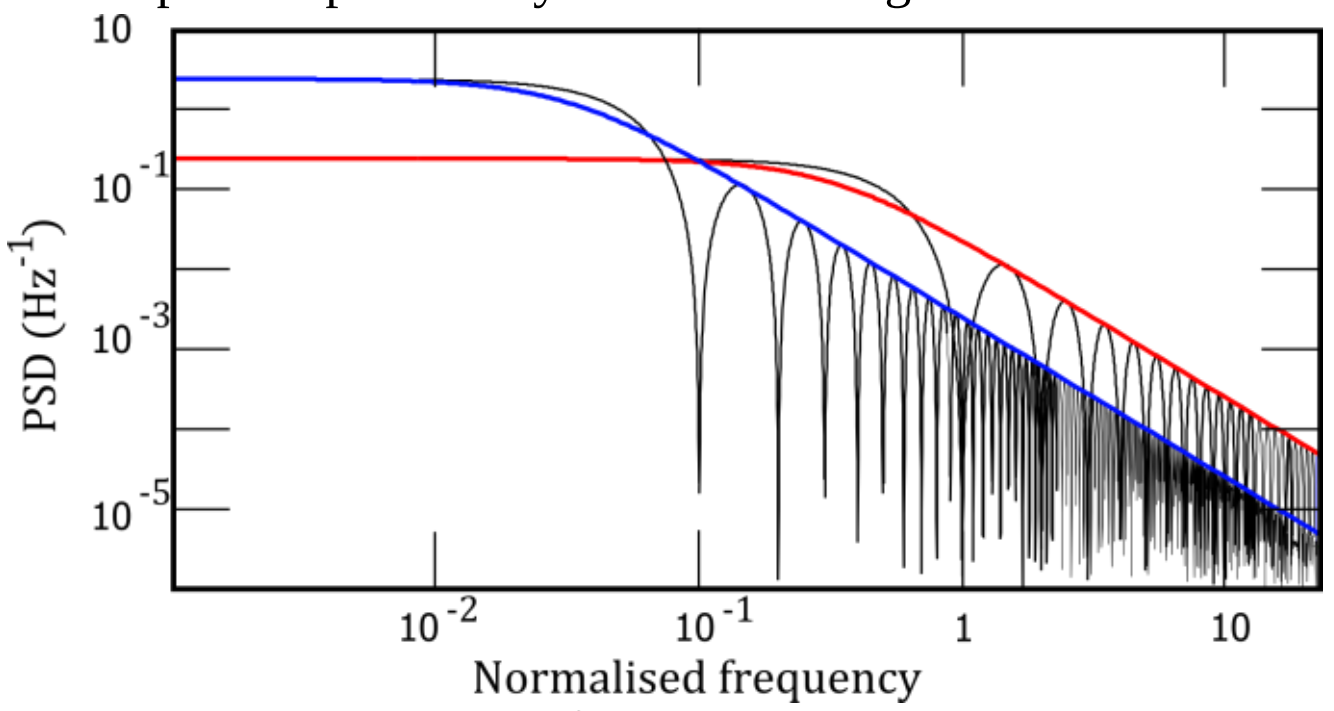

Fig. 3: Comparing the envelope functions (red and blue) with their RS PSDs (black). It can be remarked that the proposed envelope function is a good approximation to each exact PSD in all regions of interest.

By using FRS, a further shaping of the power spectrum can occur. Depicted as the red solid line in Fig. 2. is a FRS scheme which randomly switches with probability $p$ and independently keeps this state for $\ell t_\epsilon$ seconds, where the integer $\ell \in [1,10]$ is chosen according a given probability distribution. By choosing the probabilities of the various values of $\ell$, it is possible to mix between the performance of the high frequency RS and the low frequency RS. When $\mathbb{P}(\ell = 1) = 1$, then the FRS scheme becomes the high frequency RS and when $\mathbb{P}(\ell = 10) = 1$, then it becomes the low frequency RS scheme. The FRS scheme can therefore interpolate between the bounds dictated by two end-most RS.

## D. Noise Envelope Functions for the RS and FRS Schemes

It is useful to be able to describe the two types of power spectral densities more simply. The desired goal is to have a simpler noise envelope function which has the correct low frequency noise level, corner frequency and high frequency asymptote. Ideally, the true PSD should be less than or equal to the envelope at all frequencies. No regard is placed on modelling the DC Dirac delta function since this is trivial.

Given that both the RS and FRS PSDs roll-off at the same rate, it is natural to have a candidate frequency domain function which is inversely proportional to $f^2$. A two parameter candidate function which is simple enough for this purpose is given by

$$\mathcal{S}_e(f) = \frac{2Gw}{w^2 + (2\pi f)^2}. \tag{42}$$

The free parameters $G$ and $w$ need to be selected and the resulting approximation tested.

Granted, this is not a systematic way of making an approximation but it is expedient and is as close to the familiar first order approximation as possible in this setting. In deterministic signals and systems, a first order approximation has the expression $y_{FO}(t) = Ge^{-wt}\theta(t)$ as its time domain representation, where $G$ is the gain and $w$ is the bandwidth. Given the constraints on the time averaged auto-correlation function, see [33], [40] for details, an equivalent first order approximation in this setting would be the auto-correlation function $y_{FO}(\tau) = Ge^{-w|\tau|}$, which has a Fourier transform given by (42). The Fourier transform of the auto-correlation function is another definition of the PSD [33].

### 1) Fitting the Gain of the Approximation

An extremely useful property of the proposed envelope function is that $\int_{-\infty}^{\infty} \mathcal{S}_e(f)\mathrm{d}f = G$ and this implies immediately that

$$G = p(1-p). \tag{43}$$

Hence both the envelope function and the noise of the true PSD have the same total PSD.

### 2) RS Parameter Fit

The low frequency asymptote is found by discovering the limit as $f \to 0$. Comparing this limit for the RS case to the envelope function yields

$$w_{RS} = \frac{2}{\ell t_\epsilon}. \tag{44}$$

Both free parameters of the envelope can therefore be fit using (43) and (44).

The comparisons of the envelope approximation and its corresponding RS PSD for $\ell = 1$ and $\ell = 10$ are depicted in Fig. 3. It can be remarked that there is very good agreement, especially at high frequencies and low frequencies. The roll-off is perfectly captured as well, no high frequency is ever above the envelope. The power spectral conservation can also be seen in this figure. Note how the slower switching frequency (blue line) has better high frequency performance at the cost of worse low frequency performance.

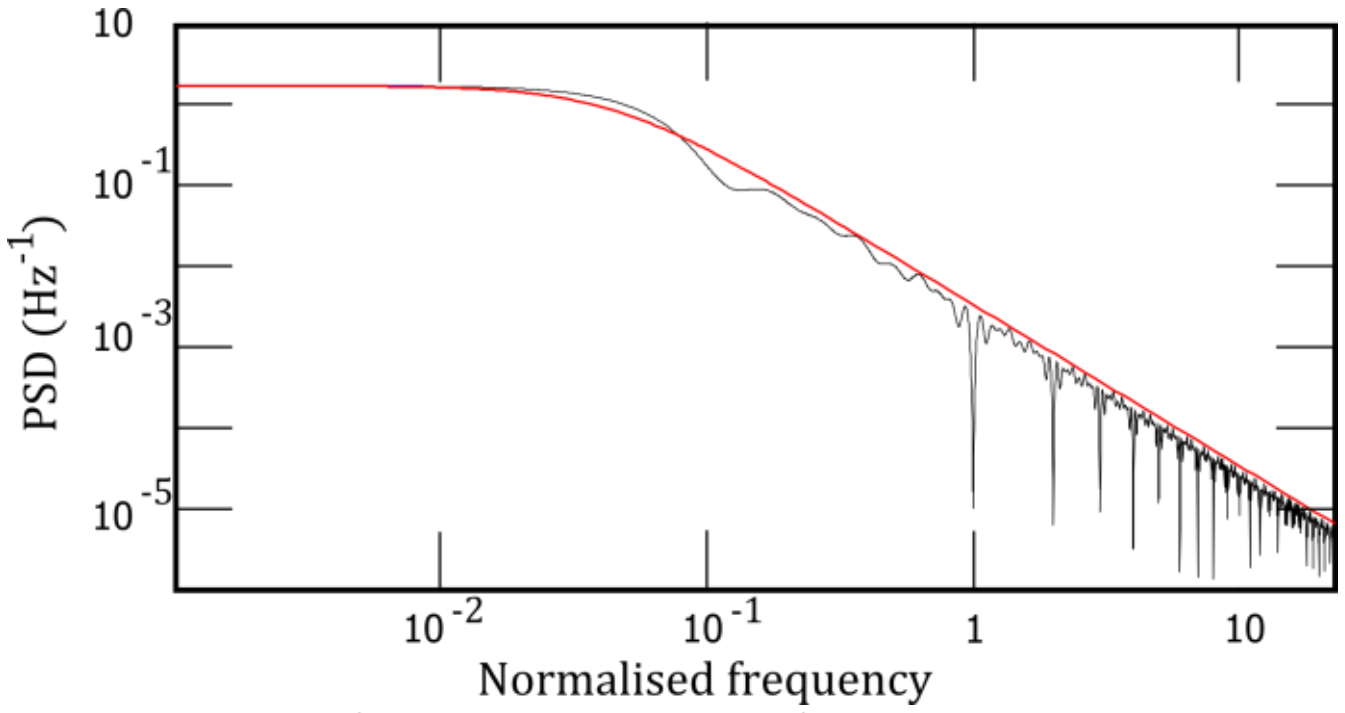

Fig. 4: Envelope function approximation for an arbitrary FRS scheme (black). Even though the probability distribution of the pulse lengths is arbitrary, it can be observed that the envelope function (red) approximates the exact analytical PSD very well. The probability distribution of the lengths is given by $\left\{\frac{9}{44}, \frac{13}{88}, 0, \frac{3}{44}, \frac{1}{44}, \frac{3}{44}, \frac{2}{11}, \frac{7}{88}, \frac{7}{44}, \frac{3}{44}\right\}$ for $\ell = \{1,2,3,4,5,6,7,8,9,10\}$.

### 3) FRS Parameter Fit

Fitting the gain of the FRS will be achieved in the exact same manner as in the RS case with the same results, namely that $G = p(1-p)$. What remains is to fit the parameter $w$ using the same technique as in the RS case.

Recall that $\bar{\ell} := \mathbb{E}_\ell\{\ell\}$ and hence discovering the low frequency limit of the FRS PSD yields

$$\lim_{f\to 0} \mathbb{E}_\ell\{|U_\ell(f)|^2\} = \mathbb{E}_\ell\{\ell^2 t_\epsilon^2\}. \tag{45}$$

Therefore, the parameter $w$ for the FRS case approximation is given by

$$w_{FRS} = \frac{2\mathbb{E}\{\ell\}}{t_\epsilon \mathbb{E}\{\ell^2\}}. \quad (46)$$

An arbitrary selection of probabilities was chosen for a ten possible length FRS scheme and it can be remarked that the envelope performs very well in this case. This is depicted in Fig. 4, where the solid red line is the envelope function and the black line is the exact analytical PSD. This kind of agreement was found to be the case upon repeated arbitrary selection of probabilities which lead us to conclude that it is correct.

It is encouraging to note that this simple approximation can faithfully model both the FRS and RS scheme's PSDs in exactly the two regions most of interest, the low frequency noise asymptote and the high frequency roll-off.

Next, semi-numerical Monte-Carlo simulations are used to validate the analytical results.

### E. Monte-Carlo Validation

In order to test the analytical results against experiment, a semi-numerical Monte-Carlo simulation was performed. It is semi-numerical in that the amplitudes (for both cases) and pulse lengths (for the FRS case) were randomly selected using numerical methods. However, the pulses to describe these were analytically manipulated using a computer algebra package. Hence, the care required when estimating the PSD using the FFT is completely avoided [2], [28]. The output of the computer algebra process is an analytical function of frequency, not a sampled data vector. An exact replica of the analytical sequence of operations needed to calculate the PSD was observed.

The limit as $T \to \infty$ was not performed and instead $N$ was made large (500 samples) and the PSD compared with the analytical predictions in (30) and (41). There is excellent agreement with the semi-numerical Monte-Carlo simulation and the exact analytical result as can be seen in Fig. 5 and Fig. 6.

The RS had a probability $p = 0.5$ and the FRS had a probability $p = 0.25$ with a uniform probability distribution of $\ell \in [1,5]$.

The simulation and theory continued to agree even when the probabilities of switching and the various possible pulse lengths were altered arbitrarily. The conclusion is therefore that the theoretical power spectral densities are the same as those calculated using this Monte-Carlo simulation.

### F. Optimum FRS

Now that the analytical results have been validated, it is an important engineering objective to be able to optimize the spectral performance of such a switching scheme. Of course, there is no such thing as a global optimum, it depends on what is trying to be achieved. One of the objectives regularly required is the minimization of discrete harmonics. Since this has already been achieved with both RS and FRS, other interests may be pursued.

As a start, the lowest possible low frequency noise level is desirable, especially in the case of DC-DC conversion. This is achieved by RS with $\mathbb{E}\{\ell\} = 1$ and $\mathbb{V}\{\ell\} = 0$. This is therefore the theoretical limit in several respects.

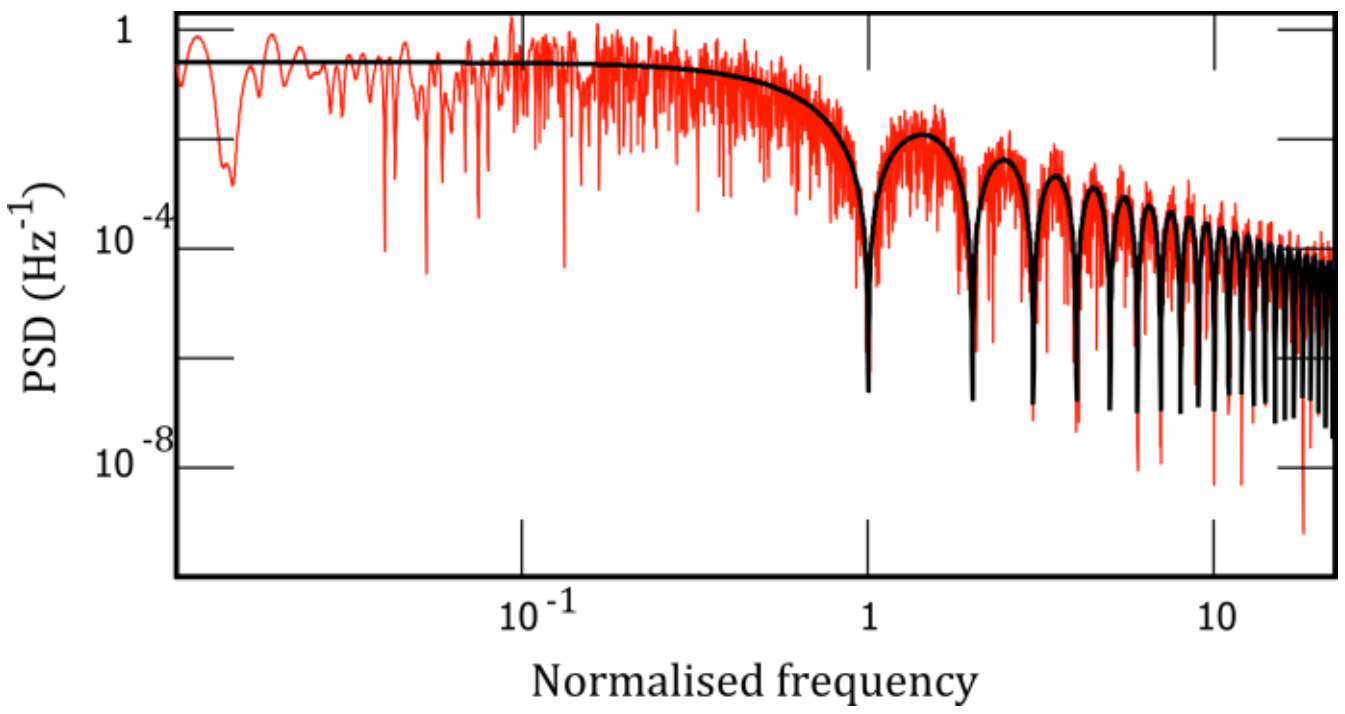

Fig. 5: PSDs of semi-numerical Monte-Carlo simulation of RS (red) and exact analytical RS (black). Visually it is apparent that with the analytical result must be correct.

In addition to spreading the noise out as much as physically possible, no other switching scheme can operate at such a high frequency and still alter the DC characteristics of the switching signal whilst simultaneously removing all discrete switching harmonics. Recall that RS in this case means switching randomly every $t_\epsilon$ seconds; which is the fastest that the switching technology can possibly switch at. As an example, modern Gallium Nitride switches have $t_\epsilon \approx 20$ nSec which is a switching frequency, $f_\epsilon \approx 50$ MHz [45].

Next, a real time controllable PSD would be desirable since it could then be shaped on demand. It will be shown that this control of the PSD is achieved by the canonical probability distribution of the pulse length. The canonical probability distribution becomes a Gaussian distribution for a large number of possible pulse lengths and $\mathbb{E}\{\ell\} \gg 1$.

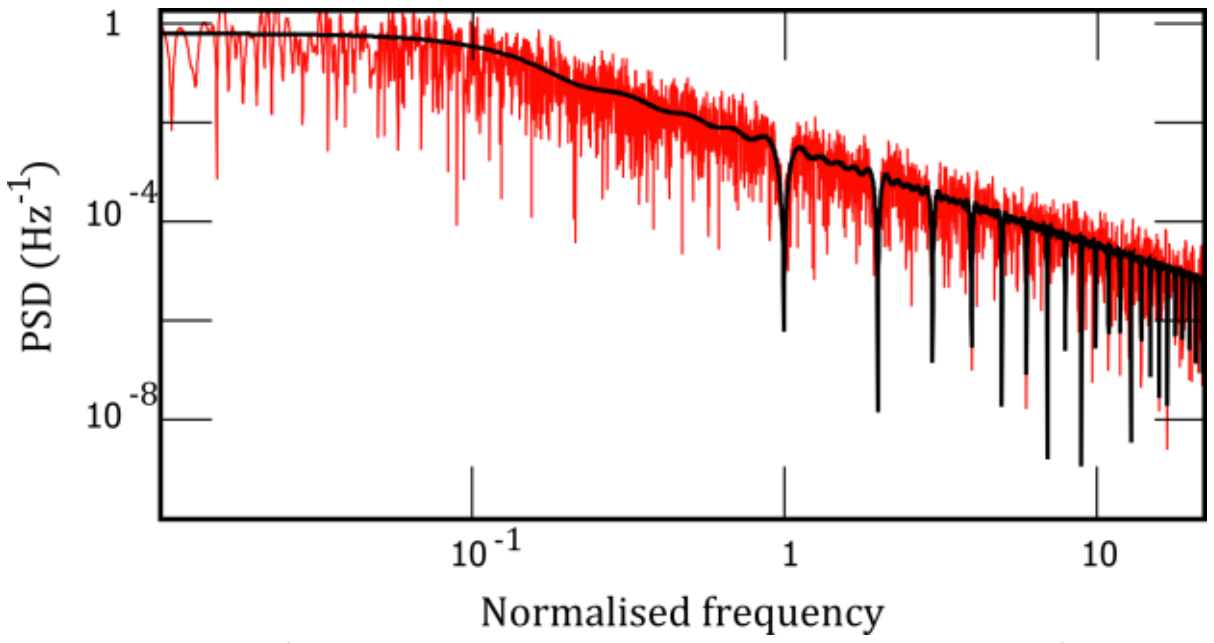

Fig. 6: PSDs of semi-numerical Monte-Carlo simulation of FRS (red) and exact analytical FRS (black). There is sufficient agreement between the analytical and simulation results that lead us to conclude the formulation is correct.

The next most desirable switching scheme would be one that results in the true noise spectrum becoming nearly indistinguishable from the envelope. This result is desirable since the peaks and valleys of the high frequency region are essentially "filled in" and better use of available spectrum is achieved. It will be shown that Huffman encoded length probabilities achieve this stated goal and yield a very good switching noise PSD. This is a special subset of the canonical distribution. In addition, the Huffman probability distribution is easy to implement digitally.

To begin the process of optimization, the low frequency noise level and high frequency noise roll-off in the switching signal are analysed first.

*1) Low Frequency Noise Level*

The low frequency noise level for the FRS scheme is given by

$$\lim_{f\to 0} \mathcal{S}_{q_\epsilon q_\epsilon}(f) - p^2\delta(f) = \lim_{f\to 0} \mathcal{S}_e(f) = \frac{2G}{w}$$
$$= p(1-p)\left(\frac{\mathbb{E}\{\ell^2\}}{\mathbb{E}\{\ell\}}\right)t_\epsilon \quad (47)$$
$$= p(1-p)\mathbb{E}\{\ell\}t_\epsilon + p(1-p)\frac{\mathbb{V}\{\ell\}}{\mathbb{E}\{\ell\}}t_\epsilon,$$

where it should be remembered that $\ell$ is an integer with a minimum value of 1. Minimising (47) is achieved trivially by making $p = 1$ or $p = 0$, in which case no switching takes place at all, or by making $\mathbb{V}\{\ell\} = 0$ and using the fact that minimum possible expected value of $\ell$ is 1. Hence the probability distribution $\mathbb{P}(\ell = 1) = 1$ achieves the minimum possible low frequency noise floor i.e. RS at a frequency $f_\epsilon$. Given this analysis and the physical impossibility of doing any other kind of switching, RS is the optimum switching scheme in many respects. It has no discrete harmonics and the PSD has been spread out as widely as possible.

However, this minimum possible low frequency noise level implies (by the conservation of PSD), the worst possible high frequency performance in the class. Quantifying the high frequency performance is turned to next.

*2) High Frequency Noise Behaviour*

The high frequency noise of the fully random PSD is difficult to analyse, hence the reason for defining the envelope. Looking at the envelope function at frequencies well above the corner frequency i.e. $f^2/w^2 > 1$, the result is that

$$\mathcal{S}_e(f) \approx \frac{2Gw}{(2\pi f)^2} = \left(\frac{4p(1-p)\mathbb{E}\{\ell\}}{\mathbb{E}\{\ell^2\}t_\epsilon}\right)\frac{1}{(2\pi f)^2}$$
$$= \frac{4p(1-p)}{t_\epsilon}\left(\frac{\mathbb{E}\{\ell\}}{\mathbb{E}\{\ell^2\}}\right)\frac{1}{(2\pi f)^2}. \quad (48)$$

Again, the mean pulse length and the variance of the pulse length are key parameters. Since these are the only two key parameters which affect the FRS scheme, it is useful to build a probability distribution which explicitly allows for their specification.

*3) The Canonical Probability Distribution*

In information theory and statistical mechanics, there is a well-known method for specifying probability distributions given only moment constraints [34], [37], [46]–[49]. It is known as the method of maximum entropy [37]. Given the constraint that $\mathbb{E}\{\ell\} = L_1$ and that $\mathbb{E}\{\ell^2\} = L_2$, the canonical probability distribution is given by

$$\mathbb{P}\{\ell\} = \frac{e^{-\alpha\ell-\beta\ell^2}}{Z} \quad (49)$$

where $Z$ is known as the partition function [37] and is calculated by

$$Z = \sum_\ell e^{-\alpha\ell-\beta\ell^2}. \quad (50)$$

Tying in the constraints is done by using the following two facts,

$$\mathbb{E}\{\ell\} = -\frac{\partial}{\partial\alpha}log\,Z\,,\ \ \mathbb{E}\{\ell^2\} = -\frac{\partial}{\partial\beta}log\,Z. \quad (51)$$

One then solves for the values of $\alpha$ and $\beta$ in (51) such that $\mathbb{E}\{\ell\} = L_1$ and $\mathbb{E}\{\ell^2\} = L_2$.

Whilst this seems simple in principle, it is not a trivial matter to solve these equations in general without resorting to numerical methods [37]. If there are a reasonably large number of possible lengths or a small variance, the discrete normal distribution is a re-parameterisation of (49) with

$$\mathbb{P}\{\ell\} = \frac{e^{\frac{-(\ell-\mu)^2}{2\nu}}}{Z} \quad (52)$$

where $Z = \sum_\ell \exp\left(\frac{-(\ell-\mu)^2}{2\nu}\right)$, $\mu = \mathbb{E}\{\ell\}$ and $\nu = \mathbb{E}\{\ell^2\} + \mu^2$

Hence, by assigning $\mu = L_1$ and $\nu = L_2 + L_1^2$, the discrete normal distribution is the family of probability distributions which can completely specify the given first and second moments without making any further unnecessary restrictions on the probability distribution.

*4) Best High Frequency and Low Frequency Trade-off*

An intuitive limiting argument will be used to arrive at the Huffman probability distribution as the best high frequency and low frequency trade-off.

Consider RS at the highest possible switching frequency, $f_\epsilon$. This has the lowest low frequency (LF) noise at the cost of the worst high frequency (HF) noise. Using RS at $f_\epsilon/2$, will alleviate the HF noise at the cost of increased LF noise. One can incorporate the benefits of the slower RS with the fastest RS by using FRS with 50% probability of either $\ell = 1$ or $\ell = 2$. This would be an optimal trade-off in the two pulse length case.

Adding in a third possible pulse length $\ell = 3$ allows for more possible sharing of the HF and LF noise. But what should be done about the pulse length probabilities to achieve this outcome? The goal is to keep both $\mathbb{E}\{\ell\}$ and $\mathbb{E}\{\ell^2\}$ as small as possible whilst including slower and slower RS.

An intuitive guess is to use a uniform probability distribution to share the benefits of all the possible RS PSDs in the expectation operator. The problem is that the limit of $\mathbb{E}\{\ell\}$ results in a divergent series i.e. $\sum_\ell \ell/N = N/2 + 1/2$ which does not have a bounded limit as $N \to \infty$.

Recursively applying the two pulse length optimal trade-off with each successive possible pulse length would be another approach. Explicitly, assign a 50% chance of $\ell = 1$ and with the remaining 50% split 25% of it with $\ell = 2$; of the remaining 25% split 12.5% of it with $\ell = 3$ and so on. There is hence a 50% chance of pulse length one, 25% chance of pulse length two etc.

The expected value of the pulse length for finite $N$ is given by $\sum_{\ell} \ell 2^{-\ell} = 2^{-N}(2^{N+1} - N - 2)$ which has a $\lim_{N\to\infty} \mathbb{E}\{\ell\} \to 2$. The second moment for finite $N$ is given by $\sum_{\ell} \ell^2 2^{-\ell} = 2^{-N}(3 \times 2^{N+1} - N^2 - 4N - 6)$ which has $\lim_{N\to\infty} \mathbb{E}\{\ell^2\} \to 6$ and therefore $\mathbb{V}\{\ell\} = 2$.

Hence, the FRS pulse length probability distribution which incorporates the HF benefits of the slower RS schemes whilst downplaying their LF failing is given by

$$\mathbb{P}\{\ell\} = \frac{2^{-\ell}}{Z}. \tag{53}$$

This probability distribution is famous in telecommunications and is an optimal variable length code [50]. The PSDs of the FRS schemes that use Huffman probabilities for their pulse lengths are depicted in Fig. 7.

The last substantial benefit of the Huffman probability distribution is that it is particularly simple to sample from. Sampling can be completed with a binary decision tree, where each branch of the tree is explored with a probability of 50%. Given the PSD of this seemingly simple sampling process, it is a recommended FRS scheme both for its HF and LF performance and ease of implementation.

TABLE I
USEFUL FRS STRATEGIES

| Design emphasis | Pulse Length Probability Distribution |
|---|---|
| Minimum low frequency noise envelope, maximal harmonic spread. | $\mathbb{P}\{\ell = 1\} = 1$ |
| Fully controllable noise PSD. | $\mathbb{P}\{\ell\} = e^{\frac{-(\ell-\mu)^2}{2v}} Z^{-1}$ |
| Best high frequency and low frequency trade-off | $\mathbb{P}\{\ell\} = 2^{-\ell} Z^{-1}$ |

*5) Useful Fully Random Probability Switching Schemes*

Table I lists a number of useful FRS schemes, each with a different design emphasis.

The theoretical limit of low frequency power spectral performance is given by RS. A "software definable" PSD filter is achieved by FRS with a Gaussian probability of pulse length. The best HF and LF trade-off is achieved by the Huffman probability distribution.

Note, selective harmonic elimination is possible with any of these schemes by choosing the minimum pulse length possible to be equal to the harmonic that should be removed. This places the zeroes of the expected sinc function at integer multiples of the minimum pulse length and zeroes out these harmonics.

## V. GENERAL DC PLUS RIPPLE MODEL: TIME DOMAIN

The following is needed to characterise the evolution of the states and the ripple of the RS and FRS switched converters. The line of reasoning followed is very similar to conventional small signal analysis for the standard DC plus ripple model [1], [51], [52]. There is one key difference; the additive signal is a random process and the ripple is not necessarily small. Let the state variable under consideration be denoted by $x(t)$ which would be either a capacitor voltage or an inductor current. The DC plus ripple model of the state variable is then

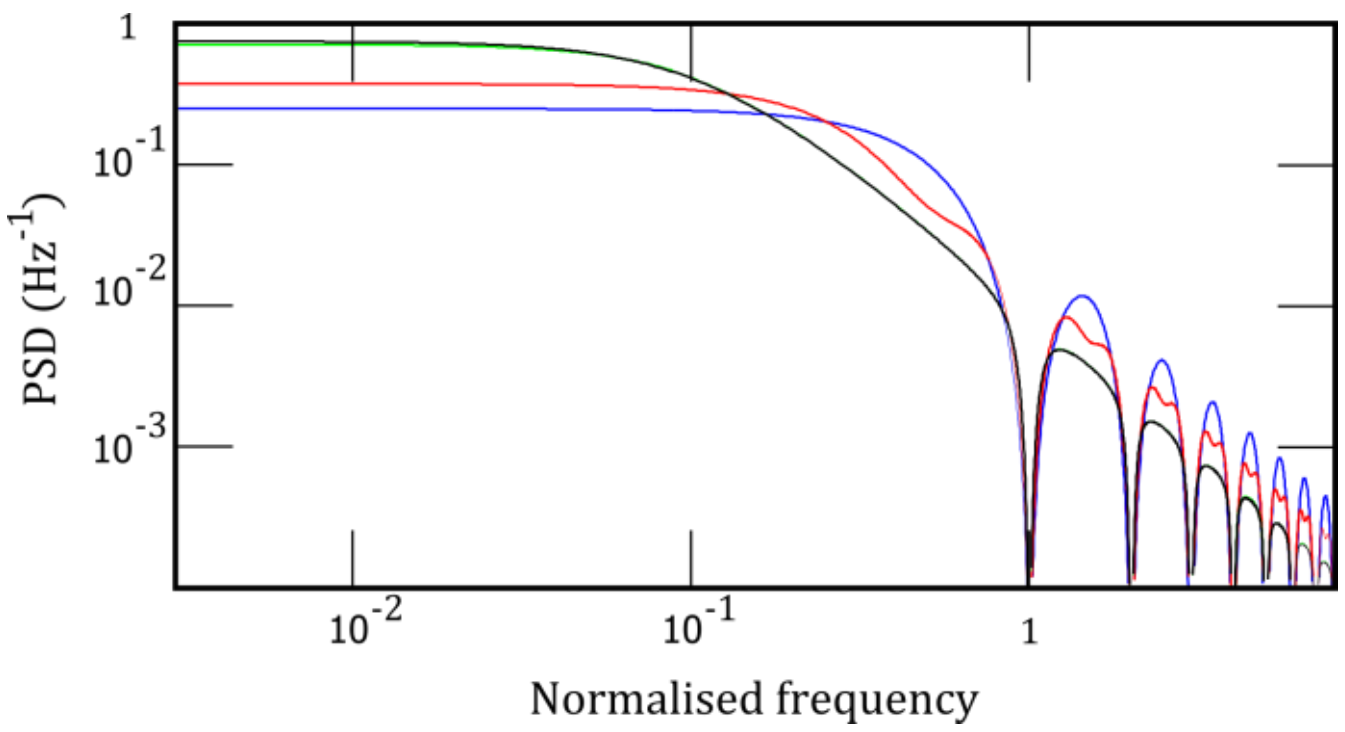

Fig. 7: PSDs of FRS with Huffman pulse length probabilities for possible pulse lengths $N$ = 1 (blue), 2 (red), 8 (green), 32 (black) and 256 (purple). Huffman FRS with more than 8 possible pulse lengths results in an indistinguishable PSD at this scale. Note that Huffman FRS with $N$ = 1 is RS at the physical limit.

given by

$$x(t) = X + \tilde{x}(t) \tag{54}$$

where

$$X \coloneqq \langle \mathbb{E}\{x(t)\} \rangle \tag{55}$$

and hence,

$$\langle \mathbb{E}\{\tilde{x}(t)\} \rangle = 0. \tag{56}$$

The proof of (56) is by definition. Consider taking the average expected value of (54), the result is

$$\begin{aligned} \langle \mathbb{E}\{x(t)\} \rangle &= X + \langle \mathbb{E}\{\tilde{x}(t)\} \rangle \\ \Rightarrow \langle \mathbb{E}\{\tilde{x}(t)\} \rangle &= \langle \mathbb{E}\{x(t)\} \rangle - X, \end{aligned} \tag{57}$$

but by definition (55) this must be

$$\langle \mathbb{E}\{\tilde{x}(t)\} \rangle = X - X = 0. \tag{58}$$

Lastly,

$$\left\langle \mathbb{E}\left\{\frac{d\tilde{x}}{dt}\right\} \right\rangle = 0. \tag{59}$$

Derivatives make additive constants zero and amplify noise, even for random processes [33]. Hence the time derivative of either a random process or a deterministic function of time which has an average expected value of zero; will have an average expected value of zero.

### A. Volt-Second and Charge Balance are Modelling Artifacts

The benefit of being explicit about these definitions up front is the following. Inductor volt-second balance and capacitor charge balance are both artifacts of the DC plus ripple model definitions. These are not vital extraneous principles for DC-DC power conversion. To see this, note the following for the case of the inductor, similar reasoning applies for the case of a capacitor.

Kirchoff's voltage law requires that,

$$v_L = L\frac{di}{dt}. \tag{60}$$

Substitution of the DC plus ripple condition implies that

$$v_L = L\frac{d}{dt}(I + \tilde{\imath}) = L\frac{d\tilde{\imath}}{dt}. \tag{61}$$

Now, taking the average expected value of both sides and using (59) means that

$$\langle \mathbb{E}\{v_L\}\rangle = 0. \tag{62}$$

Hence, the principle of inductor volt-second balance and capacitor charge balance is an artifact of DC plus ripple modeling. It is therefore not a requirement of DC-DC power conversion that it be upheld for any given length of time, it is automatically upheld provided that the circuit does not destroy itself.

### *B. Exact DC plus Ripple Switching Model*

Given a switching function of time $q(t)$, which may be deterministic or random, consider any two-configuration DC-DC converter which can be described by

$$\frac{d\boldsymbol{x}}{dt} = \left(\boldsymbol{A}_1\boldsymbol{x} + \boldsymbol{B}_1 V_g\right)q(t) + \left(\boldsymbol{A}_2\boldsymbol{x} + \boldsymbol{B}_2 V_g\right)q'(t) \tag{63}$$

Note that $\mathbf{x}$ is the vector of circuit state variables (inductor currents and capacitor voltages), the matrix $\mathbf{A}$ describes the dynamics of the system and the vector $\mathbf{B}$ describes the way the line voltage $V_g$ enters the system. Equation (63) is a non-linear, multi-variate ordinary differential equation and it is cumbersome to analytically solve it exactly.

When $q(t)$ is a PWM signal, the approximate solution to it is well known and widely applied [1]. The familiar small-ripple, or linear ripple, approximation yields a particularly simple result known as state-space averaging and there are multiple methods and assumptions which arrive at that exact same final form [1], [51], [52].

To keep the results general, the average expected value of the switching function is denoted as,

$$\langle \mathbb{E}\{q(t)\}\rangle \coloneqq p. \tag{64}$$

Even though it has been denoted $p$, the expected average of the switching function does not need to be a probability. If deterministic PWM switching is to used, then replace $p \to D$ where $D$ is the duty cycle. For RCFVD, the result would be $p \to \overline{D}$, the average duty cycle. In the case of RS though, the expected average of the switching function is indeed a probability $p$. For brevity of notation's sake, let the complement switching state $p' \coloneqq (1-p)$.

After substituting in the DC plus ripple for $\mathbf{x} \to \mathbf{X} + \tilde{\mathbf{x}}(t)$ and $q \to p + \tilde{q}(t)$ into (63), the following factorized form is arrived at,

$$\frac{d\tilde{\boldsymbol{x}}}{dt} = (p\boldsymbol{A}_1 + p'\boldsymbol{A}_2)\boldsymbol{X} + (p\boldsymbol{B}_1 + p'\boldsymbol{B}_2)V_g + \tag{65}$$
$$\left(p\boldsymbol{A}_1 + p'\boldsymbol{A}_2 + (\boldsymbol{A}_1 - \boldsymbol{A}_2)\tilde{q}(t)\right)\tilde{\boldsymbol{x}} +$$
$$\left((\boldsymbol{A}_1 - \boldsymbol{A}_2)\boldsymbol{X} + (\boldsymbol{B}_1 - \boldsymbol{B}_2)V_g\right)\tilde{q}(t)$$

Note that this is an exact large signal description of the DC-DC power converter. No assumptions or approximations have been made yet as to the nature of the ripple. Linearization has not been used nor any small-signal techniques. It is not even assumed that the switching is fast, relative to the system dynamics.

### *C. DC Solution*

Observe that by taking the expected average of both sides, the result is that

$$\begin{aligned}\left\langle \mathbb{E}\left\{\frac{d\tilde{\boldsymbol{x}}}{dt}\right\}\right\rangle &= (p\boldsymbol{A}_1 + p'\boldsymbol{A}_2)\boldsymbol{X} + (p\boldsymbol{B}_1 + p'\boldsymbol{B}_2)V_g \\ &+(p\boldsymbol{A}_1 + p'\,\boldsymbol{A}_2)\langle \mathbb{E}\{\tilde{\boldsymbol{x}}\}\rangle \\ &+(\boldsymbol{A}_1 - \boldsymbol{A}_2)\langle \mathbb{E}\{\tilde{q}(t)\tilde{\boldsymbol{x}}\}\rangle \\ &+\left((\boldsymbol{A}_1 - \boldsymbol{A}_2)\boldsymbol{X} + (\boldsymbol{B}_1 - \boldsymbol{B}_2)V_g\right)\langle \mathbb{E}\{\tilde{q}(t)\}\rangle,\end{aligned} \tag{66}$$

where the average expected value of constants left as is and linearity of the average expected value has been used. Using the facts about the average expected value of the ripple being zero, the result is that

$$0 = (p\boldsymbol{A}_1 + p'\boldsymbol{A}_2)\boldsymbol{X} + (p\boldsymbol{B}_1 + p'\boldsymbol{B}_2)V_g \tag{67}$$

which is identical to the steady state solution of state-space averaging and the small-signal AC model [1], [51], [52]. It requires some work to show that $\langle \mathbb{E}\{\tilde{q}\tilde{\mathbf{x}}\}\rangle = 0$ in general. A sketch of the reasoning is as follows: $\langle \mathbb{E}\{\tilde{x}\}\rangle = \langle \mathbb{E}\{\tilde{q}\}\rangle = 0$ by definition; also by definition $\langle \mathbb{E}\{\tilde{x}\tilde{q}\}\rangle = \rho_{\tilde{x}\tilde{q}}\sigma_{\tilde{x}}\sigma_{\tilde{q}}$, where $\rho_{\tilde{x}\tilde{q}}$ is the correlation coefficient, $\sigma_{\tilde{x}}$ is the standard deviation of the state variable and $\sigma_{\tilde{q}} = \sqrt{p(1-p)}$ is the switching ripple standard deviation. Since neither of the two standard deviations are zero, in general; the result $\langle \mathbb{E}\{\tilde{q}\tilde{\mathbf{x}}\}\rangle = 0$ can only occur if the correlation coefficient between the switching ripple and every state variable is zero. Since the switch is responsible for causing the evolution of the states, the correlation between $\tilde{q}$ and $\mathrm{d}\tilde{x}/\mathrm{d}t$ will always be $\pm 1$, depending on whether the switch causes a build-up or release of energy. The correlation between $\tilde{x}$ and $\mathrm{d}\tilde{x}/\mathrm{d}t$ is always zero and since $\mathrm{d}\tilde{x}/\mathrm{d}t$ and $\tilde{q}$ are perfectly correlated, the final implication is that the correlation between $\tilde{q}$ and $\tilde{x}$ is always zero, for every state variable.

The average state is therefore given by

$$\boldsymbol{X} = -(p\boldsymbol{A}_1 + p'\boldsymbol{A}_2)^{-1}(p\boldsymbol{B}_1 + p'\boldsymbol{B}_2)V_g. \tag{68}$$

This result is completely general, no small signal approximations were made and it is not dependent on exactly what switching scheme is implemented. As long as $\langle \mathbb{E}\{q(t)\}\rangle = p$, then (68) will calculate the DC values of the states. No assessment as to the stability of these states has been made yet, hence the ripple dynamics are looked at next.

### *D. Exact Analytical Solution for the Ripple*

To calculate the switching ripple, by definition, the average expected value of the switch is subtracted from $q(t)$. The result is that $\tilde{q} \coloneqq (1-p)$ when $q=1$ and $\tilde{q} \coloneqq -p$ when $q=0$. This fact and the substitution of the DC value from (68) into (65) means that

$$\frac{d\tilde{\boldsymbol{x}}}{dt} = \left(p\boldsymbol{A}_1 + p'\boldsymbol{A}_2 + (\boldsymbol{A}_1-\boldsymbol{A}_2)\tilde{q}(t)\right)\tilde{\boldsymbol{x}} + \left((\boldsymbol{A}_1-\boldsymbol{A}_2)\boldsymbol{X} + (\boldsymbol{B}_1-\boldsymbol{B}_2)V_g\right)\tilde{q}(t). \tag{69}$$

Hence, (70) is a switched model of the ripple dynamics,

$$\frac{d\tilde{\boldsymbol{x}}}{dt} = \begin{cases} \boldsymbol{A}_1\tilde{\boldsymbol{x}} + p'\boldsymbol{\beta}V_g, & q=1 \\ \boldsymbol{A}_2\tilde{\boldsymbol{x}} - p\,\boldsymbol{\beta}V_g, & q=0 \end{cases} \tag{70}$$

where $\boldsymbol{\beta} \coloneqq (\mathbf{A}_1-\mathbf{A}_2)\mathbf{X} + (\mathbf{B}_1-\mathbf{B}_2)$. Again, this is an exact large signal model of the ripple and it applies on an instant by instant basis depending on the value of the switching signal. It is clear that the ripple is intimately dependent on the expected average of the switch, through $p$, and the expected average state $\mathbf{X}$ which is included in $\boldsymbol{\beta}$. This result is because the second order non-linear terms which are normally ignored in the small-signal AC model were retained for this analysis. It also applies for any switching scheme.

For a given initial condition of the ripple, $\tilde{\mathbf{x}}(0)$, when $q=1$, the evolution of the ripple is given by

$$\tilde{\boldsymbol{x}}(t) = e^{\boldsymbol{A}_1 t}\tilde{\boldsymbol{x}}(0) + p'\boldsymbol{A}_1^{-1}\left(e^{\boldsymbol{A}_1 t} - \boldsymbol{I}_N\right)\boldsymbol{\beta}\,V_g \tag{71}$$

whereas when $q=0$, it is calculated by

$$\tilde{\boldsymbol{x}}(t) = e^{\boldsymbol{A}_2 t}\tilde{\boldsymbol{x}}(0) - p\boldsymbol{A}_2^{-1}\left(e^{\boldsymbol{A}_2 t} - \boldsymbol{I}_N\right)\boldsymbol{\beta}\,V_g. \tag{72}$$

Note that $e^{\boldsymbol{A}t}$ is the matrix exponential and $\mathbf{I}_\mathrm{N}$ is the $\mathrm{N} \times \mathrm{N}$ identity matrix.

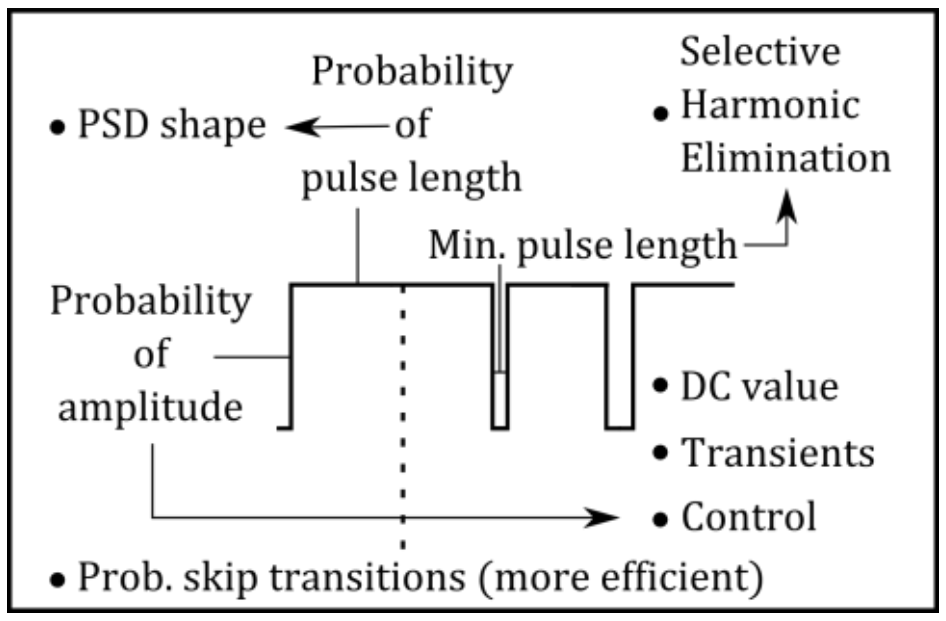

Fig. 8: A visual mnemonic to demonstrate the effect of each family of probability distributions on the circuit behavior. The probability of the amplitude affects the DC value, shapes the transient modes and allows for control. The pulse length probabilities shape the steady state PSD. The minimum pulse length results in zeroes at integer multiples of this harmonic. The switching scheme also probably skips transitions, avoiding those switching losses.

### *E. Random Switching in the Time Domain*

Using the time domain method from [53], the probability density of the evolution of the ripple may be solved for as a function of time when $q$ is randomly switched or fully randomly switched. This section will highlight an expedient means of calculating the most important aspect of the probability density for the DC-DC conversion problem, namely the expected value.

#### *1) RS Expected Value Update*

By using the linear ripple approximation and using the mean update equation from [53], in the limit as the switching frequency approaches infinity the expected value of the states update exactly the same as Cúk's state space averaging. All the salient features of the proof follow.

Assume that the pulse length time of the RS is given by a fixed value $T$ and that $\mathbf{x}_k \coloneqq \mathbf{x}(kT)$, the linear ripple approximation of the mean update equation is therefore given by

$$\mathbb{E}\{\boldsymbol{x}_{k+1}\} = \mathbb{E}\{\boldsymbol{x}_k\} + Tp\left(\boldsymbol{A}_1\mathbb{E}\{\boldsymbol{x}_k\} + \boldsymbol{B}_1V_g\right) + Tp'\left(\boldsymbol{A}_1\mathbb{E}\{\boldsymbol{x}_k\} + \boldsymbol{B}_1V_g\right) \tag{73}$$

which implies that

$$\frac{\mathbb{E}\{\boldsymbol{x}_{k+1}\} - \mathbb{E}\{\boldsymbol{x}_k\}}{T} = (p\boldsymbol{A}_1 + p'\boldsymbol{A}_2)\mathbb{E}\{\boldsymbol{x}_k\} + (p\boldsymbol{B}_1 + p'\boldsymbol{B}_2)V_g. \tag{74}$$

Taking the limit $T \to 0$ means that

$$\frac{d\mathbb{E}\{\boldsymbol{x}\}}{dt} = (p\boldsymbol{A}_1 + p'\boldsymbol{A}_2)\mathbb{E}\{\boldsymbol{x}\} + (p\boldsymbol{B}_1 + p'\boldsymbol{B}_2)V_g \tag{75}$$

which the reader will recognize as Cúk's state space averaging equation with no duty cycle or line voltage variation. Hence, the expected value dynamics are described by Cúk's state space averaging but with the change of $p \to D$. Recall that this approximation is only valid under very fast switching.

#### *2) FRS Expected Value Update Equation*

The salient features of the derivation are presented here. The difference between the RS case and the FRS case is that the value of $T$ is no longer fixed but varies probabilistically from pulse to pulse. The mean update equation can have the pulse length time marginalized out and this coupled with the linear ripple approximation results in,

$$\sum_T \mathbb{P}(T)\,\mathbb{E}\{\boldsymbol{x}_{k+1}\} = \sum_T \mathbb{P}(T)\,\mathbb{E}\{\boldsymbol{x}_k\} + \sum_T \mathbb{P}(T)\,Tp\left(\boldsymbol{A}_1\mathbb{E}\{\boldsymbol{x}_k\} + \boldsymbol{B}_1V_g\right) + \sum_T \mathbb{P}(T)\,Tp'\left(\boldsymbol{A}_1\mathbb{E}\{\boldsymbol{x}_k\} + \boldsymbol{B}_1V_g\right). \tag{76}$$

Note that $\sum_T \mathbb{P}\{T\}\,\mathbb{E}\{\mathbf{x}_{k+1}\} = \mathbb{E}\{\mathbf{x}_{k+1}\}$ and similarly for $\mathbb{E}\{\mathbf{x}_k\}$. The linearity of the right hand side of (76) then means that for FRS, the expected value of the states update as

$$\mathbb{E}\{\boldsymbol{x}_{k+1}\} = \mathbb{E}\{\boldsymbol{x}_k\} + \bar{T}p\left(\boldsymbol{A}_1\mathbb{E}\{\boldsymbol{x}_k\} + \boldsymbol{B}_1 V_g\right) + \bar{T}p'\left(\boldsymbol{A}_1\mathbb{E}\{\boldsymbol{x}_k\} + \boldsymbol{B}_1 V_g\right) \quad (77)$$

where $\bar{T} = \mathbb{E}\{T\}$. This solution is valid provided that every possible pulse length in the set results in a valid linear ripple approximation. Taking the limit as $\bar{T} \to 0$ in (106) results in the identical dynamics as described by (75). In the limit, there is no difference between the average value dynamics of RS and FRS (and PWM switching) of DC-DC power converters.

### 3) *RS and FRS Equilibrium Ripple Value*

Using the same simplification steps as above, under very fast switching, taking the ensemble average of (70) results in

$$\mathbb{E}\left\{\frac{d\tilde{\boldsymbol{x}}}{dt}\right\} = p \times \left(\boldsymbol{A}_1\tilde{\boldsymbol{x}} + p'\boldsymbol{\beta}V_g\right) + p' \times \left(\boldsymbol{A}_2\tilde{\boldsymbol{x}} - p\,\boldsymbol{\beta}V_g\right) \quad (78)$$

$$\mathbb{E}\left\{\frac{d\tilde{\boldsymbol{x}}}{dt}\right\} = (p\boldsymbol{A}_1 + p'\boldsymbol{A}_2)\mathbb{E}\{\tilde{\boldsymbol{x}}\}.$$

Taking the time average of (78), results in

$$\left\langle \mathbb{E}\left\{\frac{d\tilde{\boldsymbol{x}}}{dt}\right\}\right\rangle = (p\boldsymbol{A}_1 + p'\boldsymbol{A}_2)\langle\mathbb{E}\{\tilde{\boldsymbol{x}}\}\rangle = 0 \quad (79)$$

due to $\langle\mathbb{E}\{\tilde{\mathbf{x}}\}\rangle = 0$. The original claim about the expected average of the derivative being zero is hence verified.

Solving (78) under very fast switching results in

$$\frac{d\mathbb{E}\{\tilde{\boldsymbol{x}}\}}{dt} = (p\boldsymbol{A}_1 + p'\boldsymbol{A}_2)\mathbb{E}\{\tilde{\boldsymbol{x}}\} \quad (80)$$

with the solution

$$\mathbb{E}\{\tilde{\boldsymbol{x}}\} = e^{(p\boldsymbol{A}_1 + p'\boldsymbol{A}_2)t}\mathbb{E}\{\tilde{\boldsymbol{x}}\}_0 \quad (81)$$

where $\mathbb{E}\{\tilde{\mathbf{x}}\}_0$ is the expected value of the ripple variables at time $t = 0$. What these last two equations show is that the expected dynamics of the ripple follow the modes given by the eigenvalues of $(p\mathbf{A}_1 + p'\mathbf{A}_2)$. Hence, assuming the state-space averaged system is stable, the expected ripple dynamics are stable and approach zero exponentially. Analysing the covariance matrix update equation from [53] under fast switching produces an identical conclusion; namely that the $\mathbb{E}\{\tilde{\mathbf{x}}\} \to 0$ exponentially fast i.e. Cúk's state space average equations describe the time domain evolution of both the RS and the FRS expected values.

### 4) *Discussion on RS and FRS in the Time Domain*

It has been shown that the expected ripple exponentially goes to zero and that the expected value of the states are modelled by the state space averaging equations in both RS and FRS.

In addition, the theory shows that, under infinitely fast RS and FRS, the actual evolution of the circuit is given by the solution to (75). The circuit will behave as a deterministic system that is a blend of the two possible circuit configurations with zero ripple, as opposed to zero *average expected* ripple.

These results are gratifying in that it vindicates the use of RS. It does not matter that volt-second balance or capacitor charge balance is not guaranteed within a given time frame; as the pulse lengths of RS (and expected pulse lengths of FRS) shorten, the system behaves more and more like a deterministic one and the instantaneous deviations away from the moving average go to zero in the limit. So in addition to being the only viable switching scheme at the lowest possible time limit $t_\epsilon$, there is no loss of design ability with regards to transients, transfer functions, impedances etc. since the standard modeling tool is applicable directly.

Since only $p$ affects the transfer functions, impedances etc. the probability of pulse length, $\ell$ can be used to independently shape the switching PSD. Effectively, one can create a filter for the harmonics using the probability of the pulse lengths. This statement is depicted in Fig. 8 as a visual mnemonic.

What follows is the calculation of the PSD of the state variables under random and FRS. This will allow for a holistic filter design methodology. The effect of the circuit's filtering components as well as probability of the pulse lengths in random and FRS will be shown.

## VI. DC PLUS RIPPLE: FREQUENCY DOMAIN

Two key aspects of random processes are looked at before presenting the general result. These are the mixing equation and the random input filtering theorem.

### 1) *The Mixing Equation: Linear Combinations and Linear Combinations of Derivatives*

The calculation of the PSD requires, in general, a squaring of a random variable. This squaring ends up multiplying each term in a sum by every other term and care is required to calculate the final form of the PSD. Two key aspects of this mixing of terms are looked at now.

For linear combinations the following result shows 'mixing', in the time domain.

Let $x(t)$ be a random process made up as a linear combination of $q(t)$ i.e.

$$x(t) \coloneqq aq(t) + b, \quad (82)$$

where $a$ and $b$ are constants. Recall that the PSD of $x$ is calculated by

$$\mathcal{S}_{xx}(f) = \mathcal{F}\{\langle\mathbb{E}\{x(t)x(t+\tau)\}\rangle\}. \quad (83)$$

Let $x(t) \coloneqq x$ and $x(t+\tau) \coloneqq x'$, then by following similar notation and multiplying out the definition of $x$, one gets

$$\mathcal{S}_{xx}(f) = \mathcal{F}\{\langle\mathbb{E}\{a^2qq' + abq + abq' + b^2\}\rangle\}. \quad (84)$$

Taking the expected average of both $abq$ and $abq'$ in the time domain before taking the Fourier transform, yields $ab\langle\mathbb{E}\{q\}\rangle\delta(f)$.

Hence, the final result is

$$\mathcal{S}_{xx}(f) = a^2\mathcal{S}_{qq}(f) + 2ab\langle\mathbb{E}\{q\}\rangle\delta(f) + b^2\delta(f). \quad (85)$$

For linear combinations of derivatives, 'mixing' in the frequency domain shows that cross-terms cancel out.

Consider that $\mathcal{S}_{yy}(f)$ is known. The goal is to describe $\mathcal{S}_{ww}(f)$ which is related to linear derivatives of $y$ with

$$w = a\frac{dy}{dt} + by. \tag{86}$$

Using the definition and 'mixing' in the frequency domain yields,

$$\mathcal{S}_{ww}(f) = \langle \mathbb{E}\{W_T(f)W_T(f)^*\}\rangle \tag{87}$$

where $W_T(f)$ is the Fourier transform of a finite length of the signal $w(t)$. Now, applying the conventional Fourier transform to the right hand side of (87) the result is

$$\begin{aligned} \mathcal{S}_{ww}(f) &= \langle \mathbb{E}\{(a\mathbb{j}\omega Y_T(f) + bY_T(f))(a\mathbb{j}\omega Y_T(f) + bY_T(f))^*\}\rangle \\ &= a^2\omega^2\langle \mathbb{E}\{Y_T(f)Y_T(f)^*\}\rangle + b^2\langle \mathbb{E}\{Y_T(f)Y_T(f)^*\}\rangle \\ &= a\omega^2\mathcal{S}_{yy}(f) + b^2\mathcal{S}_{yy}(f) \end{aligned} \tag{88}$$

because the cross terms cancelled out due to the complex conjugation. This is what is meant by the mixing equation, linear combinations and linear combinations of derivatives must be mixed by multiplying through all of the terms and the combined PSD is calculated via this process. Note: if the PSD of $w$ were known and $y$ were the required PSD, then the random input filtering theorem would need to be used. This is described next.

*2) Random Filtering Theorem*

Consider passing a random signal $q(t)$ as an input to a linear time invariant system with output $y(t)$ and impulse response $h(t)$. The PSD of the output variable will be calculated by

$$\mathcal{S}_{yy}(f) = |H|^2\mathcal{S}_{qq}(f) \tag{89}$$

where $H(f)$ is the transfer function of the system. This is a well-known result and the details of the proof are in [33].

*3) Switching Ripple PSD*

By using the definition of the switching ripple, $\tilde{q} = q - \langle \mathbb{E}\{q\}\rangle$ and the linear combination theorem, the ripple PSD of the switch is given by

$$\mathcal{S}_{\tilde{q}\tilde{q}}(f) = \mathcal{S}_{qq}(f) - p^2\delta(f) \tag{90}$$

which is useful for this application.

*4) Ripple PSD*

The DC plus ripple equation is capable of solving for both the DC values of the state variables and the exact ripple of the state variables. By making the usual approximation, namely that the cross term $\tilde{q}\tilde{\mathbf{x}}$ is negligible, the result is that the ripple is described by

$$\frac{d\tilde{\boldsymbol{x}}}{dt} = (p\boldsymbol{A}_1 + p'\boldsymbol{A}_2)\tilde{\boldsymbol{x}} + \boldsymbol{\beta}\tilde{q}(t) \tag{91}$$

where all of the terms have been previously defined. The transfer function from the switching ripple, $\tilde{q}$ to the state variable ripple $\tilde{\mathbf{x}}$ is given by

$$\boldsymbol{H}(s) = \left(s\boldsymbol{I}_N - (p\boldsymbol{A}_1 + p'\boldsymbol{A}_2)\right)^{-1}\boldsymbol{\beta}. \tag{92}$$

It is crucial to note that this transfer function is identical to the control to state transfer function from Cúk's state space averaging method. This means that the filtering of the RS function is achieved via a well-known transfer function without having to resort to any expected value integrals i.e. the random input filtering theorem applies directly. It can also be thought of as the *best linear approximation* to the actual non-linear system response [54]. Hence the vector of circuit ripple power spectral densities is given by

$$\begin{bmatrix} \mathcal{S}_{\widetilde{x_1x_1}}(f) \\ \mathcal{S}_{\widetilde{x_2x_2}}(f) \\ \vdots \\ \mathcal{S}_{\widetilde{x_Nx_N}}(f) \end{bmatrix} = |\boldsymbol{H}(\mathbb{j}\omega)|^2\mathcal{S}_{\tilde{q}\tilde{q}}(f). \tag{93}$$

This is an expeditious means of calculating the influence of the RS scheme on the ripple power spectral densities. Indeed, given the general character of the random input filtering theorem, this result can be repurposed for arbitrary switching schemes, including deterministic switching [33].

Calculating the expected squared amplitude of the ripple can be achieved via the Parseval-Plancheral theorem with

$$\begin{bmatrix} \langle \mathbb{E}\{\tilde{x}_1^2\}\rangle \\ \langle \mathbb{E}\{\tilde{x}_2^2\}\rangle \\ \vdots \\ \langle \mathbb{E}\{\tilde{x}_N^2\}\rangle \end{bmatrix} = \int_{-\infty}^{\infty} |\boldsymbol{H}(\mathbb{j}\omega)|^2\mathcal{S}_{\tilde{q}\tilde{q}}(f)df, \tag{94}$$

or may be accomplished algebraically using the equilibrium covariance matrix method in [53]. The diagonal of the equilibrium covariance matrix gives each state-variable's ripple expected squared amplitude. Taking the square root of the result has the same units as the DC value and is equal to the standard deviation of the state variable of interest. In more familiar terminology, this is the RMS error of the state variable. The mathematical definition of the RMS error, also known as the standard deviation, is given by

$$\begin{aligned} \sigma_x &:= \sqrt{\langle \mathbb{E}\{\tilde{x}^2\}\rangle} \\ &= \sqrt{\langle \mathbb{E}\{x^2\}\rangle - X^2}. \end{aligned} \tag{95}$$

At steady-state, the randomly switched converter has state variables best described by a multi-variate Gaussian distribution with mean $\mathbf{X}$ and covariance matrix calculated by [53]. The amplitude histogram of each state variable is therefore expected to be normally distributed.

The previous exposition can be summarized by following these four steps.

1. Write the circuit equations for the state variables, multiplied appropriately by the switching function $q(t)$.
2. Substitute in the DC plus ripple condition and solve for the DC operating point for all circuit variables by taking the expected average, $\langle \mathbb{E}\{\ \}\rangle$, on both sides of the equation.
3. Solve for the switching to state transfer function. Using this transfer function, compute the ripple PSD of all circuit variables.
4. Calculate the RMS error of the state variables using Parseval-Plancheral's theorem OR the equilibrium covariance matrix method from [53].

Since the DC values of all of the circuit variables are already known from step 2, the PSD of all circuit variables will be written as

$$\mathcal{S}_{xx}(f) = \mathcal{S}_{\tilde{x}\tilde{x}}(f) + X^2\delta(f), \tag{96}$$

where the DC value $X$ and the ripple PSD $\mathcal{S}_{\tilde{x}\tilde{x}}(f)$ are discovered from steps 2 and 3 respectively.

*B. Example: Simplified Buck Converter*

Consider the ideal Buck converter with a resistance $r$ in series with the inductor to model conductive losses. Let $i$ be the inductor current and $v$ be the capacitor voltage. For reasons of brevity, explicitly showing the dependence on time of all functions has been suppressed.

**Step 1** involves finding the circuit equations, these are given by

$$L\frac{di}{dt} = q(t)\left(V_g - ir\right) - v \qquad C\frac{dv}{dt} = i - \frac{v}{R} \tag{97}$$

**Step 2** involves substituting in the DC plus ripple condition, hence

$$L\frac{d\tilde{\imath}}{dt} = \left(p + \tilde{q}(t)\right)\left(V_g - (I + \tilde{\imath})r\right) - (V + \tilde{v}) \qquad C\frac{d\tilde{v}}{dt} = I + \tilde{\imath} - \frac{(V + \tilde{v})}{R}, \tag{98}$$

and taking $\langle \mathbb{E}\{\ \}\rangle$ on both sides, one gets

$$0 = pV_g - pIr - V \qquad 0 = I - \frac{V}{R}. \tag{99}$$

Hence

$$V = \frac{pV_gR}{(R + pr)} \qquad I = \frac{pV_g}{(R + pr)} \tag{100}$$

which is identical to the usual circuit average steady state solution with $D \to p$ as previously declared.

**Step 3** Replacing the DC values from (99) into the ripple model of (98) (assuming $\tilde{q}\tilde{\mathbf{x}}$ is negligible) results in

$$L\frac{d\tilde{\imath}}{dt} + pr\tilde{\imath} + \tilde{v} = V_g\alpha\tilde{q} \qquad C\frac{d\tilde{v}}{dt} + \frac{\tilde{v}}{R} = \tilde{\imath} \tag{101}$$

where $\alpha = \frac{R}{R+pr}$.

Hence, the transfer function can be found using matrix algebra at this stage. However, this problem is simple enough to do by hand and show the steps.

Decoupling this differential equation may be done by inserting the capacitor current equation and its derivative into the inductor voltage equation, hence

$$LC\frac{d^2\tilde{v}}{dt^2} + \left(\frac{L}{R} + rpC\right)\frac{d\tilde{v}}{dt} + \tilde{v}\left(1 + \frac{pr}{R}\right) = V_g\alpha\tilde{q}. \tag{102}$$

This is a second order LTI system being driven by a single random input $q(t)$, hence the PSD of the output ripple voltage is given by

$$\mathcal{S}_{\tilde{v}\tilde{v}}(f) = \left|\frac{1}{s^2LC + s\left(\frac{L}{R} + rpC\right) + \left(1 + p\frac{r}{R}\right)}\right|^2 V_g^2\alpha^2\mathcal{S}_{\tilde{q}\tilde{q}}(f)\,, \tag{103}$$

where $s = \mathbb{j}2\pi f$. Hence, the ripple voltage PSD is given by

$$\mathcal{S}_{\tilde{v}\tilde{v}}(f) = |H(f)|^2V_g^2\alpha^2\mathcal{S}_{\tilde{q}\tilde{q}}(f). \tag{104}$$

Calculating the current PSD can be accomplished by the linear combination of derivatives. Using (101) and the fact that $\mathcal{S}_{\tilde{v}\tilde{v}}(f)$ is known results in

$$\mathcal{S}_{\tilde{\imath}\tilde{\imath}}(f) = \left(C^2\omega^2 + \frac{1}{R^2}\right)\mathcal{S}_{\tilde{v}\tilde{v}}(f) = \frac{(R^2C^2\omega^2 + 1)|H(f)|^2V_g^2\alpha^2}{R^2}\mathcal{S}_{\tilde{q}\tilde{q}}(f). \tag{105}$$

One can check that this is the same small signal AC transfer function which would result from a perturbation of the duty cycle $d$ to $i$ in the conventional theory [1].

**Step 4** involves finding the RMS error of the voltage and current. There are many ways to compute these integrals. The reason for the envelope approximation of the switching ripple was to be able to approximate the switching ripple PSD for just this kind of purpose. Alternatively, the equilibrium covariance matrix method may be used. Using the equilibrium covariance matrix method (assuming linear ripple) and a computer algebra package the result is (106), where $f_s$ is the switching frequency, $\nu = RC(R + pr) + L$ and $\gamma = 2prC^2LR^3 - 2f_s^{-1}CLR^3 - pr^2f_s^{-1}C^2R^3 + 2CL^2R^2 + 2p^2r^2C^2LR^2 - 6prf_s^{-1}CLR^2 - p^2r^3f_s^{-1}C^2R^2 + 2prCL^2R - f_s^{-1}L^2R - 3p^2r^2f_s^{-1}CLR - pr^2f_s^{-1}CLR - prf_s^{-1}L^2$.

$$\sigma_i^2 = RC\frac{p(1-p)V_g^2}{f_s}\left(\frac{R}{(R+pr)}\right)^2\frac{\nu}{\gamma}$$
$$\sigma_v^2 = \frac{\sigma_i^2}{\nu}LR^3 \tag{106}$$

*1) Discussion: Noise and Heat*

An important and non-trivial result is found by considering the effect that the series resistance $r$ has on the ripple power spectral densities and the DC values.

Using the standard methodology, it can be shown that the average expected input power is equal to

$$\langle\mathbb{E}\{P_{in}\}\rangle = V_g pI$$
$$= \frac{V_g^2\,p^2}{(R+pr)} \tag{107}$$

and that the average expected output power is approximately given by

$$\langle\mathbb{E}\{P_{out}\}\rangle \approx \frac{V^2}{R}$$
$$= \frac{p^2V_g^2R}{(R+pr)^2} \tag{108}$$

since the output power harmonics are ignored in the standard framework. The efficiency is therefore approximated by

$$\eta = \frac{\langle\mathbb{E}\{P_{out}\}\rangle}{\langle\mathbb{E}\{P_{in}\}\rangle}$$
$$= \frac{R}{(R+pr)}. \tag{109}$$

In the limit as $r \to 0$, the efficiency approaches unity since there are no conduction losses.

This efficiency calculation is useful since the term appears directly in the noise PSD, it is exactly equal to $\alpha$, the gain of the ripple LTI system i.e.

$$\mathcal{S}_{\widetilde{v}\widetilde{v}}(f)$$
$$= \eta^2\left|\frac{1}{LCs^2+\left(\frac{L}{R}+rpC\right)s+\left(1+\frac{pr}{R}\right)}\right|^2 V_g^2\mathcal{S}_{\widetilde{q}\widetilde{q}}(f).$$

(1

where $s = \mathbb{j}2\pi f$. Looking at the low frequency asymptote of the output voltage noise yields the following,

$$\lim_{f\to 0}\mathcal{S}_{\widetilde{v}\widetilde{v}}(f) = \frac{\eta^2}{\left(1+\frac{pr}{R}\right)}V_g^2\mathcal{S}_{\widetilde{q}\widetilde{q}}(0)$$
$$= \eta V_g^2 p(1-p)t_\epsilon\left(\mathbb{E}\{\ell\}+\frac{\mathbb{V}\{\ell\}}{\mathbb{E}\{\ell\}}\right). \tag{111}$$

It is a given that the RS parameter, $p$ sets the DC behavior of the device. It cannot be altered in order to achieve a better PSD. The fundamental switching period $t_\epsilon$ is a given fixed value and can also not be altered. The only parameters which can alter the spectral performance of the buck converter under RS and FRS are therefore $\mathbb{E}\{\ell\}$ and $\mathbb{V}\{\ell\}$ and $\eta$. Operating right at the theoretical limit means FRS with $\mathbb{E}\{\ell\} = 1$ and $\mathbb{V}\{\ell\} = 0$, hence

$$min(max\,\mathcal{S}_{\widetilde{v}\widetilde{v}}(f)) = \eta V_g^2 p(1-p)t_\epsilon. \tag{112}$$

This is the best voltage noise ceiling possible for a switched-mode buck converter without resorting to an EMI filter. It has the noise harmonics spread as widely as possible in the frequency domain, has zero discrete harmonics and is fully controllable right at the fastest possible switching frequency. Of note is that the theoretical limit has a noise ceiling which is intimately dependent on the efficiency of the device. One is able to reduce the low frequency noise even further only at the expense of efficiency being reduced. The noise has to be transformed into heat in order to reduce it any further in other words. It can therefore be postulated that this is the true purpose of the EMI filter, it transforms the excess noise into heat. Note that the noise power has units of $\text{Volt}^2/\text{Hz}$.

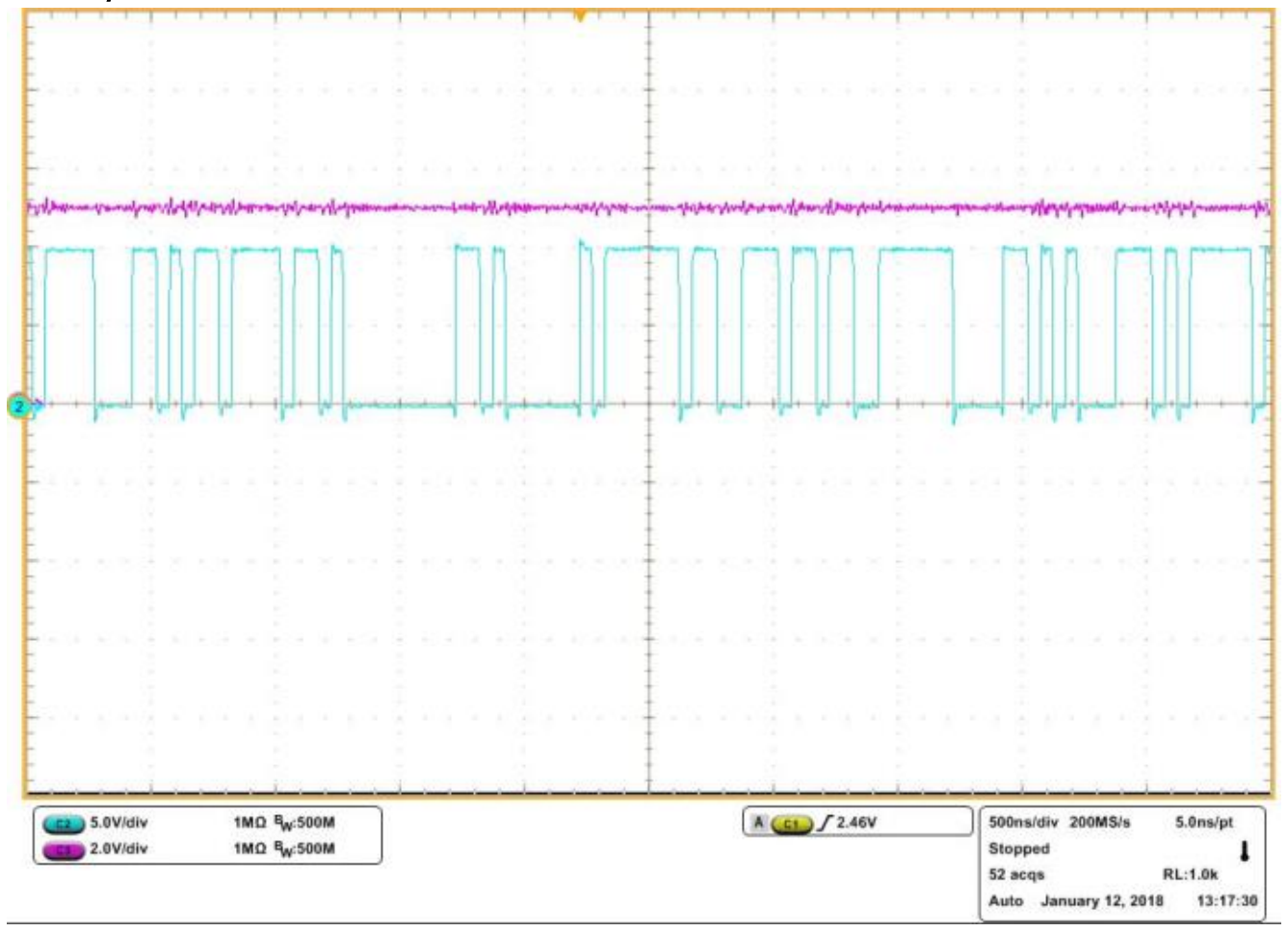

Fig. 9: Output voltage of a buck converter due to RS scheme. Note the extended periods of time which have zero switching transitions as predicted.

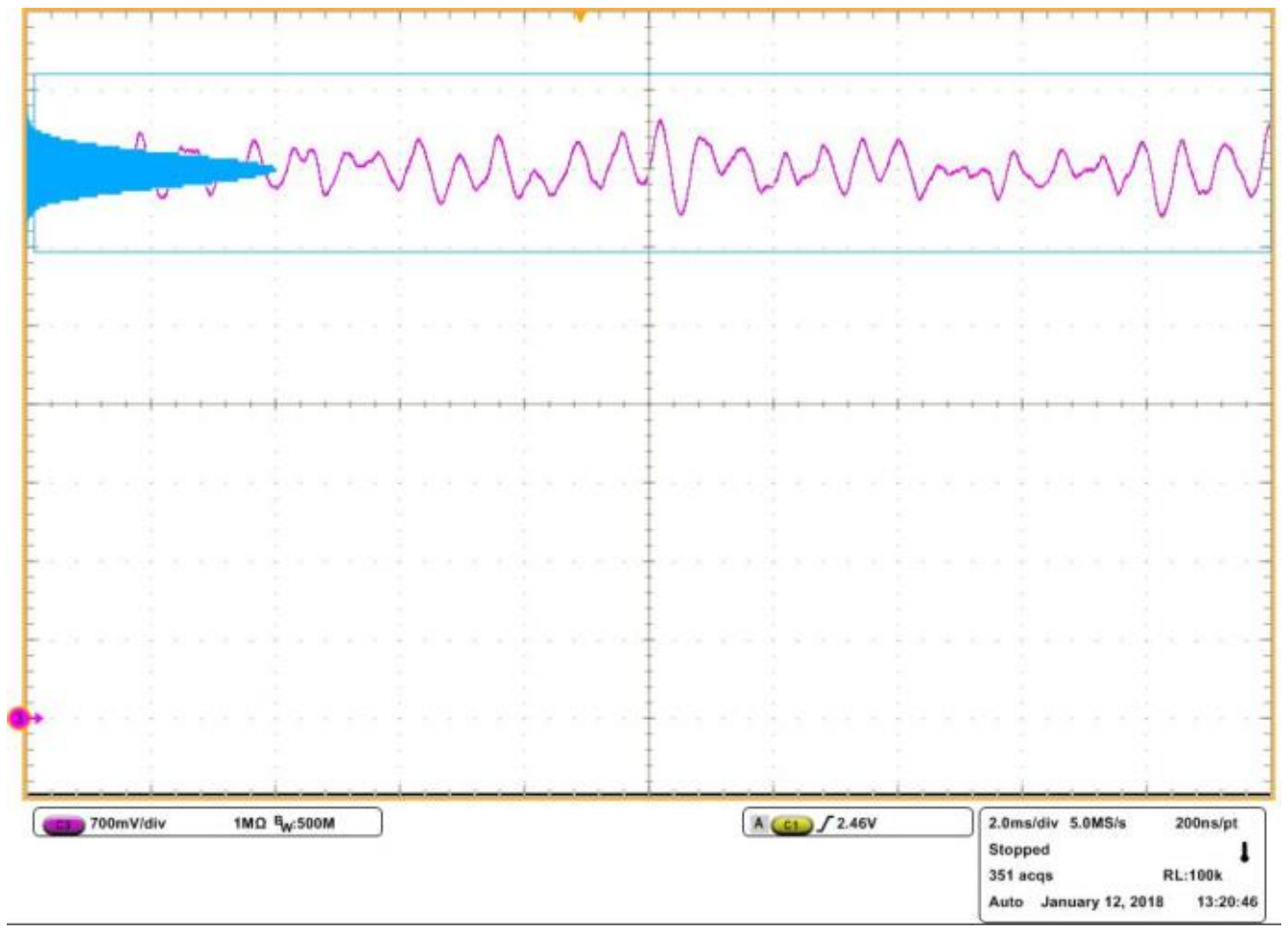

Fig. 10: High resolution version of Fig. 9 with histogram. Note that the output voltage is normally distributed as predicted by the theory of RS.

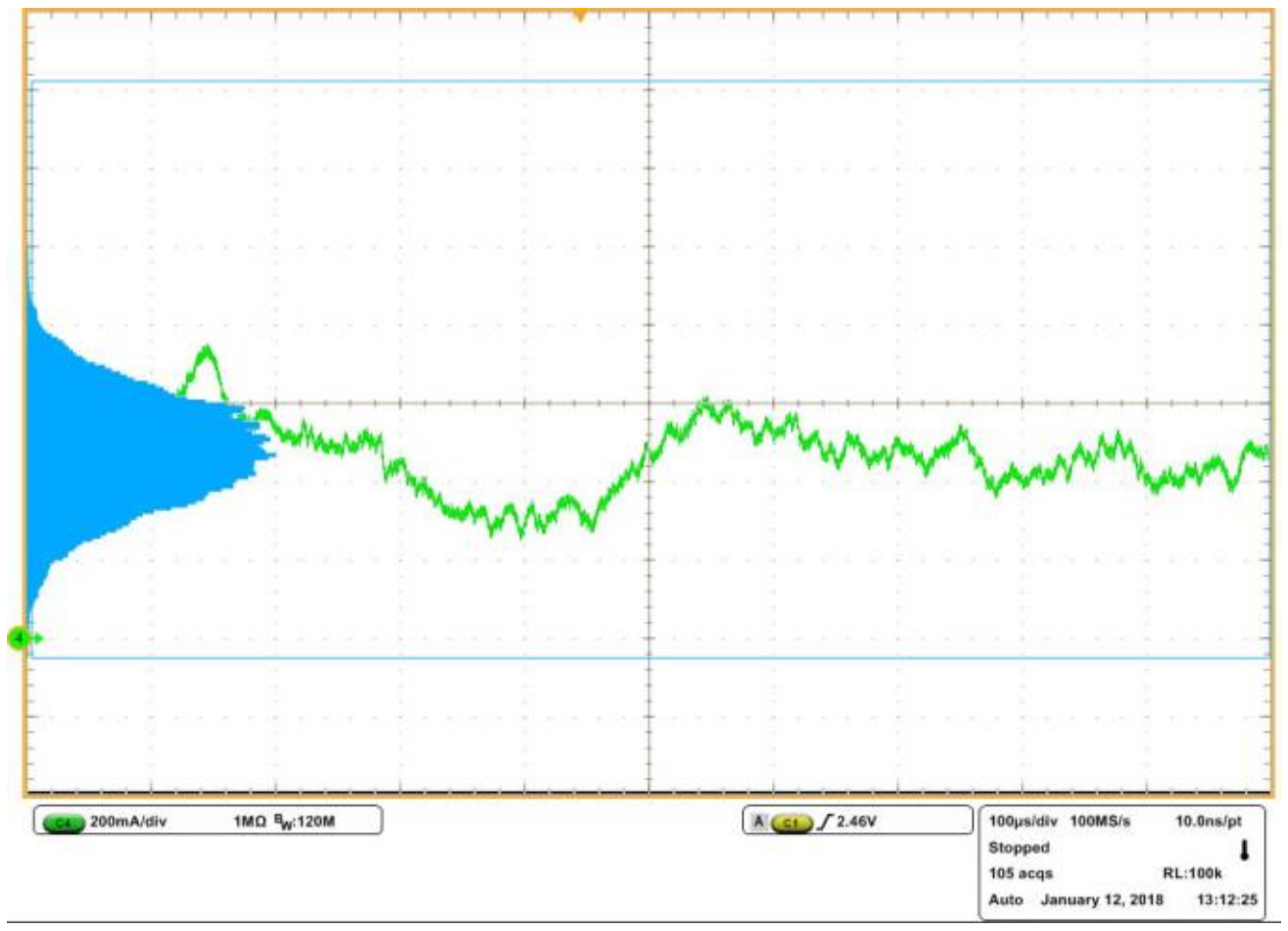

Fig. 11: High resolution oscilloscope reading of the inductor current of the buck converter. The current is normally distributed as expected.

### *C. Experimental Verification*

A 5 V buck converter fed from a 10 V source was developed to demonstrate the veracity of these analytical claims. A 32-bit Linear Feedback Shift Register (LFSR) pseudo-random number generator, driven by a 20 MHz clock was used to feed the gate drivers for the GaN switches on an EPC 9006 half-bridge demo board. The output filter consisted of a 101 μH inductor and 100 μF capacitor with a load of 10 Ω. The results of the experiment are depicted in Fig. 9, 10 and 11. Using the variance calculations from (106), with $r = 0$ as a first approximation, the theoretical results are that $\sigma_i = 249.39$ mA and $\sigma_v = 249.38$ mV.

## VII. Random Switching: control aspects

Since Cúk's state space averaging model applies directly to the RS and FRS schemes. Control of the device being driven by these switching schemes can therefore be achieved using the standard tool. A few unique features with regards to RS and FRS control will be looked at in this section. Quasi-static random control, whereby the switching probability $p$ is slowly altered as a function of time will be explored. There are theoretical performance limits with regards to this strategy which are described in detail.

Three important closed loop control strategies will be looked at, RS with Hysteresis, Random Integral Control, and Random State Feedback control. One of the major benefits of these kinds of closed loop switching schemes is that no filtering is required in the feedback loop. As an intuition primer, consider that a closed loop RS scheme will randomly respond to the output variable(s), by its very nature of operation. Only when it responds to errors as often as it fails to respond to errors will the system remain steady. It is shown that this steady state case is exactly the equilibrium condition sought. Therefore no filtering is required in control the DC value, only a biased random response to error.

### *A. Quasi-Static Control*

Quasi-static control works by ensuring that the time variation of $p$ is so slow that the system has time to reach equilibrium before the next change in value of $p$, it was inspired from the work in classical thermodynamics [55], [56]. This type of control was previously described in [57]. Recall that the state-space averaged system's modes of decay also depend on $p$, so the definition of "slow" changes as the value of $p$ changes. This would be classified as an open loop control strategy since no feedback is utilized and an explicit model of the plant is required in order to calculate the appropriate value of $p$ to arrive at a given DC value of the system states $\mathbf{X}$.

Using either random or FRS, the DC value of the switching function is made to be a slow function of time $p \to p(t)$. It is not assumed that the DC variation is small, only that it is slow. Hence, the equilibrium states are now also a slow function of time $\mathbf{X} \to \mathbf{X}(t)$ and these are approximately calculated by

$$\boldsymbol{X}(t) \approx (p(t)\boldsymbol{A}_1 + p'(t)\boldsymbol{A}_2)^{-1}(p(t)\boldsymbol{B}_1 + p'(t)\boldsymbol{B}_2)V_g. \quad (113)$$

The definition of slow depends on the proposed value of $p$ in the next step. Perturb the present value of $p \to p + \Delta p$, where $\Delta p$ is not necessarily small. The new equilibrium state variable is denoted as $\mathbf{X}'$ whereas the present one is denoted by $\mathbf{X}$. Provided that the system is being switched fast enough so that linear ripple applies at the micro-time scale, the expected value of the ripple states (error) will decay with dynamics governed by (81) i.e.

$$\mathbb{E}\{\widetilde{\boldsymbol{x}}\} = e^{\left(p\boldsymbol{A}_1 + p'\boldsymbol{A}_2 + \Delta p(\boldsymbol{A}_1 - \boldsymbol{A}_2)\right)t}\Delta\boldsymbol{X} \quad (114)$$

where $\Delta\mathbf{X} = \mathbf{X}' - \mathbf{X}$. Hence, the definition of slow is the time it takes $\mathbb{E}\{\tilde{\mathbf{x}}\} \approx 0$. This time would be dominated by the slowest eigenvalue of the argument of the exponential decay in (106). The most expedient means of calculating this slowest eigenvalue is by using a matrix algebra theorem that relates the 2-norm of the inverse of a matrix to the smallest eigenvalue [43]. This means that the ripple will decay approximately with

$$\mathbb{E}\{\widetilde{\boldsymbol{x}}\} \approx e^{-\lambda_{min}t}\Delta\boldsymbol{X} \quad (115)$$

where

$$\lambda_{min} = -\mathcal{R}e\left\{\frac{1}{\left\|\left(p\boldsymbol{A}_1 + p'\boldsymbol{A}_2 + \Delta p(\boldsymbol{A}_1 - \boldsymbol{A}_2)\right)^{-1}\right\|_2}\right\}. \quad (116)$$

So provided that $t > \frac{5}{\lambda_{\min}}$, the change induced by $\Delta p$ will be within 1% of the final equilibrium value i.e.

$$\left|\frac{\Delta p}{\Delta t}\right| < \frac{\lambda_{min}}{5}. \quad (117)$$

Hence, (117) is the formal definition of slow such that quasi-static control is possible.

### *B. Closed Loop Control*

Let $\mathbf{X}_d$ be the desired DC value of the states of the converter. This means that the desired reference states are given by $\mathbf{x}_d(t) = \mathbf{X}_d$ since the desired DC condition has no ripple. The errors between the actual states and the desired states are therefore given by the vector

$$\boldsymbol{e}(t) = \boldsymbol{x}_d(t) - \boldsymbol{x}(t) \\ = \boldsymbol{X}_d - \boldsymbol{X} - \widetilde{\boldsymbol{x}}(t). \tag{118}$$

Hence, provided that $\mathbf{X} = \mathbf{X}_d$, the ripple vector completely defines the time domain description of the errors. Hence, the ripple dynamics describe the evolution of the error in a DC-DC converter. The noise PSD is therefore the error PSD and the RMS error is a justified measure of the total ripple.

Recall the property that the probability of the amplitude in both random and FRS specifies completely the DC behavior whereas the probability of the pulse lengths specify the noise spectrum. Hence, the control of the randomly driven DC-DC converter can be achieved entirely by consideration of the probability of the amplitude only. In an open loop configuration, (4) describes the amplitudes of both RS and FRS. In closed loop, the probability of the amplitude being equal to 1 is conditional on some function of the measurement of the state(s)

$$\mathbb{P}\{a_k = 1 |\, m(\boldsymbol{x})\} = sat\left(p_{ref} + f\big(m(\boldsymbol{x})\big)\right), \tag{119}$$

where $p_{\text{ref}}$ is a reference probability, $m(\mathbf{x})$ is a measurement of the process, $f(\quad)$ is the feedback function and $\text{sat}(\quad)$ is the saturation function which ensures that the probability is bounded between 0 and 1. The saturation function is defined as

$$sat(u) \coloneqq \begin{cases} 1, & u \geq 1 \\ 0, & u \leq 0. \end{cases} \tag{120}$$

By the rules of probability theory, since probability has to sum to unity, the probability of the amplitude being equal to 0 is given by

$$\mathbb{P}\{a_k = 0 |\, m(\boldsymbol{x})\} = \\ 1 - sat\left(p_{ref} + f\big(m(\boldsymbol{x})\big)\right). \tag{121}$$

Any other choice would violate a fundamental rule of probability theory. Hence, the conditional expected value of the switching amplitude at any instant in time $kt_\epsilon$ is therefore

$$\mathbb{E}\{a_k | m(\boldsymbol{x}_k)\} = sat\left(p_{ref} + f\big(m(\boldsymbol{x}_k)\big)\right) \times 1 \\ + \\ \left(1 - sat\left(p_{ref} - f\big(m(\boldsymbol{x}_k)\big)\right)\right) \times 0 \\ = sat\left(p_{ref} + f\big(m(\boldsymbol{x}_k)\big)\right). \tag{122}$$

Any RS control algorithm will therefore be defined as a choice of $p_{\text{ref}}$ and $f\big(m(\mathbf{x}_k)\big)$.

## C. RS with Hysteresis

RS with hysteresis randomly switches with a reference probability such that the DC value will be correct; but it has a defined operational envelope such that any state which exceeds a threshold triggers the switch in order to bring that state back under control. Essentially this is RS with safety limits which limit the maximum possible random drifting of the variables. As an example, the reference probability could be 50% so that a buck converter halves the input voltage whilst the hysteresis toggles the switch whenever the current is too high or too low. This example would have a start-up sequence whereby the switch latches in the “on” state to drive the current towards equilibrium and once it clears the lower threshold, begins randomly switching at 50% probability. If at any point, either due to random chance, load changes etc. the current exceeds a threshold, it is toggled to bring the current back within the hysteresis band. This type of strategy is depicted in Fig. 12 in order to give a clearer understanding.

The first step of designing this kind of control scheme involves choosing $p_{\text{ref}}$ such that the DC value of the states are the desired ones. Mathematically this means solving for $p_{\text{ref}}$ such that

$$\boldsymbol{X} = \left(p_{ref}\boldsymbol{A}_1 + p'_{ref}\boldsymbol{A}_2\right)^{-1}\big(p_{ref}\boldsymbol{B}_1 \\ + p'_{ref}\boldsymbol{B}_2\big)V_g = \boldsymbol{X}_d \tag{123}$$

If more than one state variable is to have this hysteresis and the threshold levels are chosen poorly, then it is not difficult to predict that limit cycles could form. For example, in response to an over voltage, the switch is toggled but the correction dynamics are such that there is an over current condition which toggles the switch again which leads to an over voltage and so on. This type of limit cycle can be avoided as long as the hysteresis bands are wide enough to allow ‘wandering’. To further this argument, consider the actual current behavior in the RS buck converter depicted in Fig. 11. If the hysteresis bands are so large that the entire probable amplitude span is included; then the system is effectively operating under open loop control. If the band is reduced to exclude certain current levels then of course the ‘tails’ of the open-loop normal distribution will be eliminated at the cost of an increase in switching events.

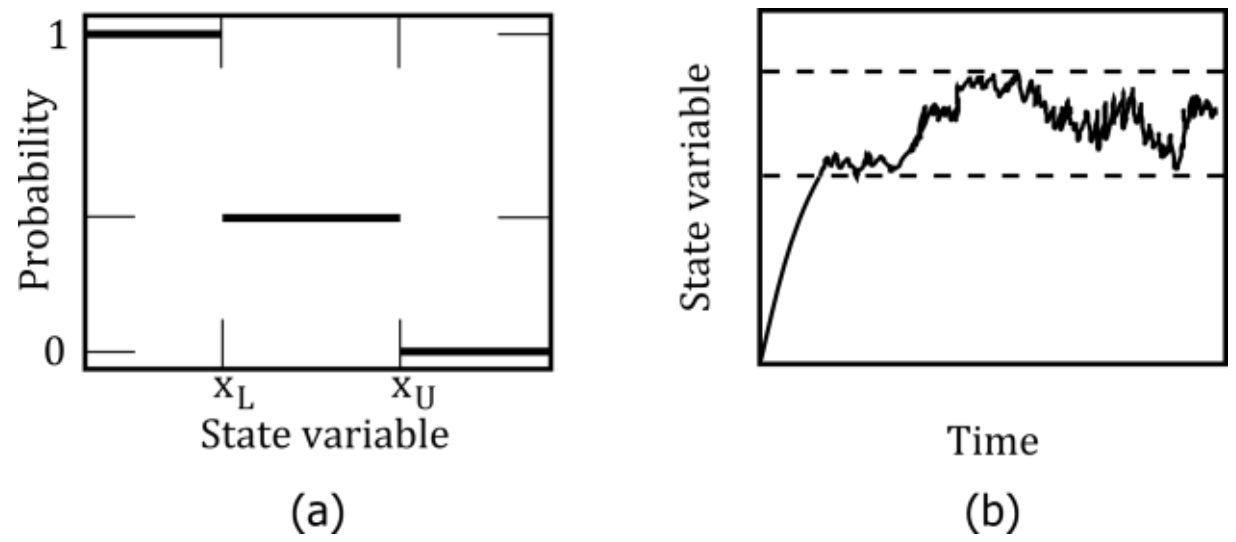

Fig. 12: (a) RS control with hysteresis. The instantaneous value of the state variable will saturate the probability outside of the hysteresis bounded by $[x_L, x_U]$. The probability of switching is a constant within the hysteresis band. (b) The salient start-up behavior of RS control with hysteresis. Note that the state variable never escapes the hysteresis band after entering it.

In all non-trivial cases, the hysteresis band will introduce additional switching events which in turn will introduce additional switching losses that were not present in the open loop case. The result is an excess heat production i.e. noise will have been traded for heat.

### D. Random Integral Control

This type of control is important because it does not rely on a model in order to determine the correct value of $p_{\text{ref}}$. The multi-variate extension of the idea requires some care and the point of the device under control is DC-DC conversion, hence the integrator is made to operate on the error of the output voltage only. Let the state of the integrator be $s_I$, the random integral control algorithm would therefore be defined by

$$\mathbb{P}\{a_k = 1|\boldsymbol{x}\} = sat(k_I s_I)$$
$$\frac{ds_I}{dt} = \big(V_d - v(t)\big). \tag{124}$$

Due to the fact that saturation is explicit in the control algorithm, there is a possibility of integrator wind-up if the integrator gain $k_I$ is chosen badly [58]. Anti-windup protection on this kind of integral control is hence important [58]. What anti-windup protection does is stop the integration of errors whenever the probability of the amplitude saturates at either 1 or 0.

The proof that integral control will eventually find the appropriate reference probability relies on the DC plus ripple model of output voltage. Using the DC plus ripple condition, the differential equation of the probabilistic integral controller is given by,

$$\frac{ds_I}{dt} = V_d - V - \tilde{v}(t) \tag{125}$$

which has an average expected value of

$$\left\langle \mathbb{E}\left\{\frac{ds_I}{dt}\right\}\right\rangle = V_d - V - \langle \mathbb{E}\{\tilde{v}(t)\}\rangle. \tag{126}$$

However, from the definitions, the average expected value of the ripple voltage is zero and the average expected rate of change of the probability of switching is therefore given by

$$\left\langle \mathbb{E}\left\{\frac{ds_I}{dt}\right\}\right\rangle = V_d - V. \tag{127}$$

Therefore, if $V \neq V_d$, the average expected rate of change of the probability of switching will increase (or decrease) depending on the offset. This increase (or decrease) will bring $V$ closer to $V_d$ and eventually the average expected rate of change of the switching probability will equal to zero. The system will therefore be in steady state with the correct DC output voltage. Given this state of affairs, at equilibrium the instantaneous time rate of change of the probability of switching will equal to

$$\frac{d\tilde{s_I}}{dt} = -\tilde{v}(t). \tag{128}$$

Hence, the switching probability as a function of time (assuming no saturation takes place) will be

$$\mathbb{P}\{a_k = 1|\boldsymbol{x}\} = k_I s_I(t)$$
$$= k_I S_I + k_I \tilde{s_I}(t) = p^* + k_I \tilde{s_I}(t) \tag{129}$$

where $p^* = k_I S_I$ is the equilibrium switching probability such that the DC output voltage is correct. The fluctuations about the correct switching probability, $\tilde{s_I}$ will be automatically filtered by a number of things. The value of $k_I$ will scale the overall level of voltage fluctuations' effect on $s_I$. The filtering of high frequency process noise will be achieved by the integrator in (128) and finally the switch will only *probably* respond to the fluctuations in $s_I$, assuming no saturation occurs. All three of these effects are beneficial in this context.

For successful probabilistic integral control, it is important that $k_I$ be small. A quantitative measure of *small* can be inferred from the quasi-static conditions. If

$$\left|\frac{ds_I}{dt}\right| \leq \frac{\lambda_{min}}{5k_I} \tag{130}$$

then the change in switching probability will change quasi-statically and the integrator will slowly find the correct switching probability such that the output DC voltage is correct.

### E. Random State Feedback

Due to Cúk's state space averaging equation applying, random state feedback control is not different from the usual PWM state feedback control other than with the change of $D \rightarrow p$. Hence the controller for random state feedback control will be given by

$$\mathbb{P}\{a_k|\boldsymbol{x}\} = sat\left(p_{ref} - \boldsymbol{K}(\boldsymbol{x}_d - \boldsymbol{x})\right) \tag{131}$$

where $p_{\text{ref}}$ is chosen such that when the error is zero, the correct DC value is maintained. As with the random switching with Hysteresis, $p_{\text{ref}}$ may be chosen by solving the DC equation directly. Another model-free way to achieve the same would be to use random integral control to calculate $p_{\text{ref}}$ on-line and use state-feedback to shape the closed loop poles so the desired transient behavior is achieved. The usual care must be taken when designing the closed loop poles such that **K** is not too large [1].

## VIII. Conclusion

This paper has demonstrated the analysis and design of RS and FRS switching schemes which offer several benefits over conventional PWM with random dithering schemes. It should be possible to shape the narrow-band performance using the zeroes of the sinc function. Paradoxically, the RS and FRS PSDs show that eliminating a narrow-band harmonic can be achieved by randomly switching at an integer multiple of that frequency. Selective harmonic elimination can be therefore be accomplished in real time very simply but only at multiples of a single frequency. Quantifying the shape of the narrow band and the how the pulse length probabilities influence the shape of the switching signals' PSD around the zeroes would therefore be a useful piece of analysis in order to further refine the analysis of this sort of selective harmonic elimination.

## Appendix

The angle bracket or averaging operator is defined by

$$\langle f(t) \rangle \coloneqq \lim_{T\to\infty} \frac{1}{2T} \int_{-T}^{T} f(t)dt \tag{132}$$

which has the following important properties.

If $a$ is a constant then,

$$\langle a \rangle = a \tag{133}$$

whereas if $f_p(t)$ is periodic with period $\tau$ then it can be proven that

$$\langle f_p(t) \rangle = \frac{1}{\tau} \int_{0}^{\tau} f_p(t) \mathrm{d}t. \tag{134}$$

Lastly, transients which decay to zero have an angle bracket which equals zero i.e.

$$\langle f_\delta(t) \rangle = 0 \tag{135}$$

where $\lim_{T\to\infty} f_\delta(T) = \lim_{T\to\infty} f_\delta(-T) \to 0$.